\documentclass[%
 reprint,
 amsmath,amssymb,
 aps,
prl,
]{revtex4-1}
\pdfoutput=1
\usepackage{graphicx}
\usepackage{dcolumn}
\usepackage{bm}
\usepackage{siunitx}
\DeclareSIUnit\gauss{G}
\usepackage{braket}
\usepackage{xcolor}
\usepackage{hyperref}

\newcommand{\vect}[1]{\mathbf{#1}}
\newcommand{\eB}[1]{\epsilon_{\rm{B},\vect{#1}}}
\newcommand{\eI}[1]{\epsilon_{\rm{I},\vect{#1}}}
\def\NB{N_{\rm B}}
\def\nB{n_{\rm B}}

\begin{document}

\title{Bose polarons near quantum criticality}

\author{Zoe Z. Yan$^{1}$, Yiqi Ni$^{1}$, Carsten Robens$^{1}$, and Martin W. Zwierlein$^{1}$}

\affiliation{
 $^{1}$MIT-Harvard Center for Ultracold Atoms, Research Laboratory of Electronics, and Department of Physics,
 Massachusetts Institute of Technology, Cambridge, Massachusetts 02139, USA
}%

\date{\today}

\begin{abstract}
The emergence of quasiparticles in strongly interacting matter represents one of the cornerstones of modern physics. However, when different phases of matter compete near a quantum critical point, the very existence of quasiparticles comes under question.
Here we create Bose polarons near quantum criticality by immersing atomic impurities in a Bose-Einstein condensate (BEC) with near-resonant interactions.
Using locally-resolved radiofrequency spectroscopy, we probe the energy, spectral width, and short-range correlations of the impurities as a function of temperature.
Far below the superfluid critical temperature, the impurities form well-defined quasiparticles.
However, their inverse lifetime, given by their spectral width, is observed to increase linearly with temperature at the \textit{Planckian} scale $\frac{k_B T}{\hbar}$, a hallmark of quantum critical behavior.
Close to the BEC critical temperature, the spectral width exceeds the binding energy of the impurities, signaling a breakdown of the quasiparticle picture.
\end{abstract}

\pacs{Valid PACS appear here}

\maketitle

%%BEGIN INTRODUCTION %%%%%%%%%%%%%%%%%%%%%%%%%%%%%%%%%%%%%%%%%%%%%%%%%%%%%%%%%%%%%%%%

	A great success of quantum many-body physics is the description of a large variety of strongly interacting systems by a collection of weakly interacting quasiparticles~\cite{Nozieres1999}.
A paradigmatic example of such a quasiparticle is an electron propagating through an ionic crystal. As anticipated by Landau~\cite{Landau1933},  Pekar found that the electron can create its own bound state by polarizing its environment~\cite{Pekar1946T,transPekar1946b}. The electron dressed by lattice distortions forms a quasiparticle, which he named the \textit{polaron}.
The polaron concept~\cite{Frohlich1954,Feynman1955} finds wide application across condensed matter physics, in phenomena ranging from
colossal magneto-resistance, to charge transport in organic semiconductors, and to high-temperature superconductivity~\cite{Alexandrov2010}.
However, near quantum phase transitions, where different phases of matter compete, the quasiparticle concept may break down~\cite{Sachdev2011}.
In such a quantum critical regime, where the temperature $T$ sets the only remaining energy scale, all relaxation times become as short as allowed by quantum mechanics, \textit{i.e.}, on the order of the \textit{Planckian} time scale $\hbar/k_B T$.
The ensuing breakdown of well-defined quasiparticles appears to be at work in the ``strange metal'' regime of cuprate superconductors, where resistivity is found to scale linearly with temperature and at the Planckian scale~\cite{Lee2006,Hartnoll2015}.
\begin{figure}[tb]
	\begin{center}
		\includegraphics[width=\columnwidth]{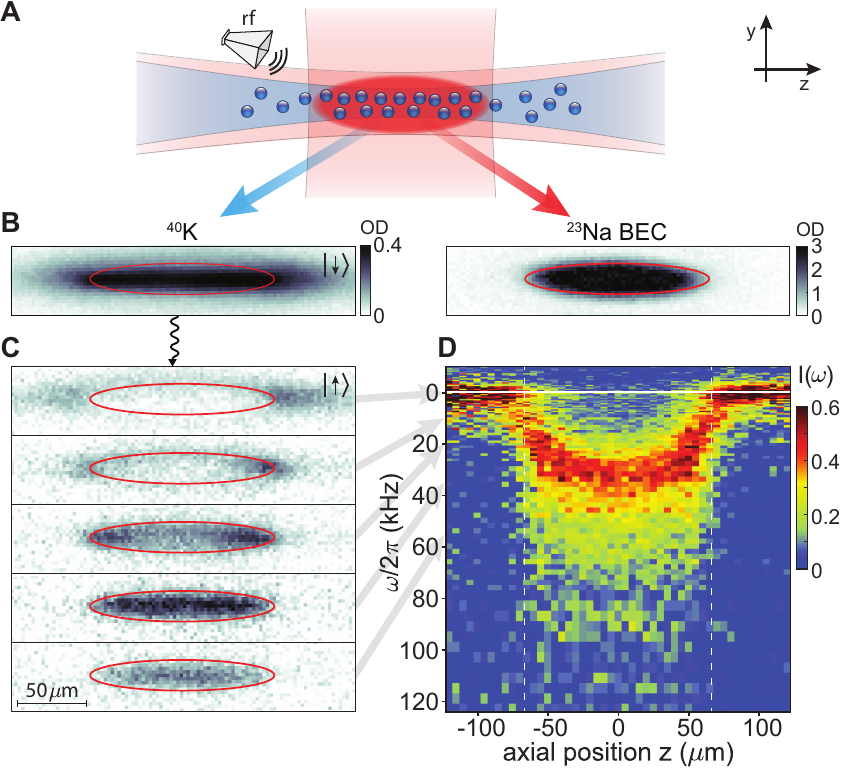}
		\caption{\label{fig:localspectrum}
			Locally-resolved radiofrequency (rf) ejection spectroscopy of strongly-coupled Bose polarons for a peak interaction strength of $(k_n a)^{-1} \,{=}\, {-}0.3$.
			\textbf{(A)} Illustration of impurities (blue) immersed in a Bose-Einstein condensate (red), both trapped in a dual-color optical dipole trap.
			\textbf{(B)} \textit{In-situ} column densities of  $^{40}$K impurities in the strongly-interacting spin state $\ket{\downarrow}$  (left) immersed in a $^{23}$Na BEC (right), where the red ellipses mark the BEC's Thomas-Fermi boundary.
			\textbf{(C)} Impurities transferred into the non-interacting $\ket{\uparrow}$ state at various rf frequencies, as indicated by the arrows.
			\textbf{(D)} Local rf transfer $I(\omega)$ of the impurity column density as a function of axial position.
			The dashed vertical lines mark the condensate's axial Thomas-Fermi radius and the solid horizontal line at $\omega/2\pi\,{=}\,0\,\rm kHz$ denotes the bare atomic transition.}
	\end{center}
\vspace{-2em}
\end{figure}

Ultracold quantum gases provide an ideal testing ground to study the fate of quasiparticles near quantum critical points. Species composition and densities, interaction strengths, and confining geometries can be controlled in a pristine fashion~\cite{Bloch2008}.
Quantum gases close to Feshbach resonances have been shown to be controlled by quantum critical points at zero temperature, separating the vacuum of a given species from the phase at finite density~\cite{Sachdev2011,Nikolic2007,Ludwig2011,Enss2012,Bertaina2013,Supplemental}.
These points control the behavior of the gas in the quantum-critical region at non-zero temperature~\cite{Sachdev2011}.
The immersion of dilute impurities into a gas of another species with resonant mutual interactions thus places the mixture in direct vicinity to the quantum critical point separating the impurity vacuum from the phase at finite impurity density~\cite{Ludwig2011}. In addition, the impurities can serve as a sensor of quantum and classical critical behavior of the host gas itself~\cite{Sachdev1999}.
The dressing of resonant impurities into quasiparticles in a cold atom environment was first observed in the case of the Fermi polaron~\cite{Schirotzek2009,Nascimbene2009,Koschorreck2012,Kohstall2012,Cetina2016,
	Scazza2017, Yan2019}~-- an atomic impurity embedded in a Fermi sea~\cite{Chevy2006,Prokofev2008,Massignan2014,Schmidt2018}.
Impurities immersed in a Bose-Einstein condensate (BEC) have been posited to form the paradigmatic Bose polarons originally considered by Pekar~\cite{Kalas2005,Cucchietti2006,Tempere2009a}.
Predicting the Bose polaron's fate upon entering the regime of strong impurity-boson interactions has proven a challenge even at zero temperature, yielding diverging results on its properties from the ground-state energy to the effective mass~\cite{Tempere2009a,Rath2013,Christensen2015,Grusdt2015,Vlietinck2015,Ardila2015,Ardila2016,Grusdt2016,Shchadilova2016b,Yoshida2018}.
The complexity of describing the strongly-coupled Bose polaron increases further at non-zero temperatures~\cite{Levinsen2017,Guenther2018}. Already for weak interactions, the decay rate of polarons has been predicted to be strongly enhanced with increasing temperature, achieving its maximal value near the BEC transition temperature of the host gas~\cite{Levinsen2017}. Near resonance, in the quantum critical regime of the boson-impurity mixture, the very existence of a well-defined quasiparticle is in question~\cite{Sachdev2011,Ludwig2011,Bertaina2013,Supplemental}.
Experimentally, evidence of Bose polaronic phenomena was observed in the expansion~\cite{Park2012} and trapping~\cite{DeSalvo2017} of fermions immersed in a BEC, through the phononic Lamb shift~\cite{Rentrop2016}, and in the dynamics of impurities~\cite{Catani2012}.
The continuum of excited states of impurities was probed in radiofrequency (rf) \textit{injection} spectroscopy~\cite{Supplemental} on Bose-Fermi mixtures~\cite{Wu2012,Hu2016} and in a two-state mixture of bosons~\cite{Jorgensen2016}, yielding evidence for polaronic energy shifts of such excitations.

Here we create and study the strongly-coupled Bose polaron in equilibrium by immersing fermionic impurities into a Bose gas near an interspecies Feshbach resonance and explore the impurity's evolution in the quantum critical regime of the Bose-Fermi mixture, including
the onset of quantum degeneracy of the bosonic bath.
Utilizing locally-resolved \textit{ejection} spectroscopy~\cite{Supplemental}, we measure the polaron's momentum-integrated spectral function~\cite{Zwerger2016}, giving access to the polaron's energy, the quasiparticle lifetime, and the strength of short-range correlations with the host bosons. These correlations are quantified by the \textit{contact}~\cite{Zwerger2016,Baym2007,Punk2007,Tan2008,Braaten2010,Zwierlein2016}, which also captures the change in the polaron energy with interaction strength.
For near-resonant interactions, we find the spectral width $\Gamma$ -- a measure of the quasiparticle decay rate~\cite{Schirotzek2009,Yan2019,Zwerger2016,Schneider2009} -- to grow linearly with temperature at the Planckian scale $k_B T/\hbar$, and $\hbar\Gamma$ to exceed the impurity's energy close to the onset of quantum degeneracy for the bosonic bath.
These properties of the spectral width are direct signatures of quantum critical behavior.

%%  BEGIN EXPERIMENTAL DESCRIPTION %%%%%%%%%%%%%%%%%%%%%%%%%%%%%%%%%%%%%%%%%%%%%%%%%%%

The experiment starts with an ultracold gas of fermionic $^{40}$K atoms immersed in a BEC of $^{23}$Na~\cite{Park2012} at a temperature of $T\,{\approx}\,\SI{130}{\nano\kelvin}$.
Both species are trapped in an optical dipole trap as ellipsoidal atom clouds in their respective hyperfine ground states ($\ket{F\,{=}\,1, m_F\,{=}\,1}$ for $^{23}$Na and $\ket{9/2,-9/2}\,{\equiv}\,\ket{\downarrow}$ for $^{40}$K).
Peak boson and fermion densities are $n_\text{Na} \,{=}\,\SI{6e13}{\per\cubic\centi\meter}$ and $n_\text{K}\,{=}\,\SI{2e11}{\per\cubic\centi\meter}$, corresponding to an impurity concentration of $0.3\%$.
\begin{figure}
	\includegraphics[width=\columnwidth]{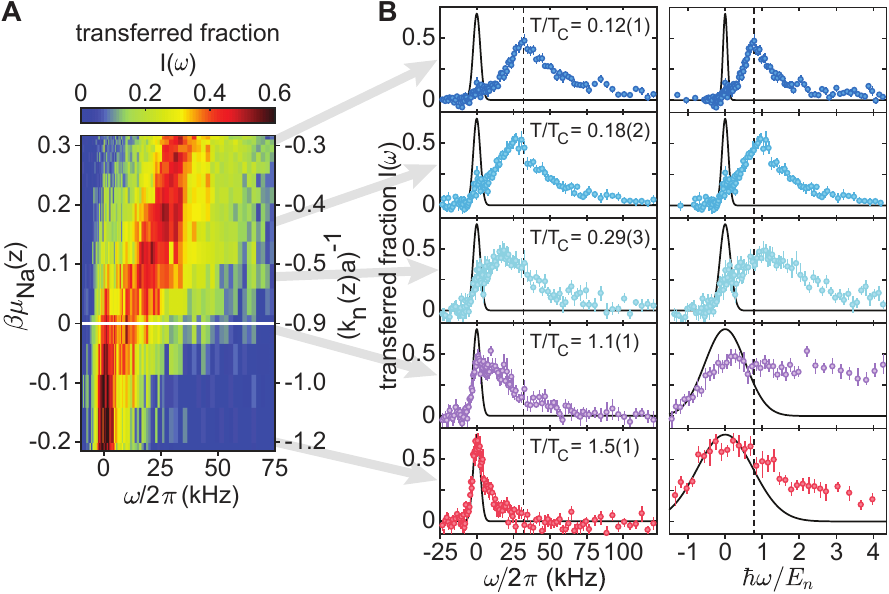}
		\caption{\label{fig:rf} Rf ejection spectra of Bose polarons at various reduced temperatures $T/T_\mathrm{C}$ with a peak interaction strength of $(k_n a)^{-1} \,{=}\, -0.3$.
	\textbf{(A)} Color density map of the rf transfer $I(\omega)$ as a function of the normalized local chemical potential $\beta\mu(z)$ and the local interaction strength $(k_na)^{-1}$.
	The solid white line marks the BEC phase transition at $\beta\mu\,{=}\,0$.
	\textbf{(B)}
	Fraction of impurities transferred into the non-interacting state $\ket{\uparrow}$ as a function of rf frequency (left) and of normalized frequency, $\hbar\omega/E_n$ (right)~\cite{Supplemental}.  The dashed black line marks the peak transfer location of the impurities at the lowest $T/T_\mathrm{C}$. The solid black lines show the rf spectrum of bare $^{40}$K atoms, indicating the spectral resolution. Error bars reflect $1\sigma$ statistical uncertainty~\cite{Supplemental}.}
\end{figure}
The BEC is weakly interacting, with an interboson scattering length of $a_{\rm BB}\,{=}\,52\,a_0$~\cite{Tiesinga1996}.
To create strongly-coupled Bose polarons in their attractive ground state, we ramp the magnetic field close to an interspecies Feshbach resonance~\cite{Park2012,Supplemental}, where impurities in the $\ket{\downarrow}$ state are strongly attracted to the sodium atoms with a peak interaction strength of $(k_n a)^{-1}\,{=}\,{-}0.3$.
Here, $k_n\,{=}\,(6\pi^2 n_\text{Na})^{1/3}\,{=}\,(1300\,a_\text{0})^{-1}$ is the inverse interboson distance, $a$ is the interspecies scattering length, and $a_\text{0}$ is the Bohr radius.
For these near-resonant interactions, the thermal equilibration time set by two-body collisions is near its unitarity-limited value of $\hbar/E_n\,{\approx}\,\SI{4}{\micro\second}$, three orders of magnitude faster than the lifetime of the gas mixture in this regime, limited by three-body losses to about $\SI{4}{\milli\second}$.
Here, $E_n\,{=}\,\hbar^2 k_n^2/4 m_{\rm r}$ is the degeneracy energy scale, and $m_{\rm r}\,{=}\,m_\text{K}m_\text{Na}/(m_\text{K}+m_\text{Na})$ is the reduced mass of the impurity-boson scattering problem.
By preparing the strongly interacting system within $\SI{2}{\milli\second}$, we can study Bose polarons in equilibrium before losses become significant.
At the chosen magnetic field, impurities in the $\ket{\downarrow}$ state are strongly interacting with the condensate, while they are non-interacting in the hyperfine state ($\ket{9/2,-7/2}\,{\equiv}\,\ket{\uparrow}$). This provides us with the ideal conditions to perform rf {ejection} spectroscopy, whereby an rf pulse transfers impurities from the interacting $\ket{\downarrow}$ state into the non-interacting $\ket{\uparrow}$ state. We employ an rf pulse of Gaussian envelope with a full-width-half-maximum resolution of \SI{6}{\kilo\hertz} and measure the fraction of impurities $I(\omega)$ transferred into the $\ket{\uparrow}$ state.

Fig.~\ref{fig:localspectrum} displays the locally-resolved rf spectrum of strongly-coupled Bose polarons.
As shown in Fig.~\ref{fig:localspectrum}A, the rf transfer $I(\omega)$ is strongly spatially dependent, and its maximum is shifted furthest from the bare atomic resonance for impurities deep inside the BEC (Fig.~\ref{fig:localspectrum}B).
Here, the rf photon must supply a significant additional amount of energy to transfer the bound impurity into the non-interacting state.
The central peak shift in Fig.~\ref{fig:localspectrum}C corresponds to an energy shift of $h{\cdot}\SI{32}{\kilo\hertz}\,{=}\,0.82\,E_n$, indicating an impurity energy that is unitarity-limited, given by the degeneracy energy scale $E_n$.
For comparison, the mean-field energy experienced by bosons in the BEC is only ${\approx}\, h {\cdot} 0.8\,\rm kHz$.
In addition to the strong shift, we also observe long tails at higher frequencies in the rf transfer, a tell-tale sign of contact interactions~\cite{Baym2007,Punk2007,Braaten2008}.

\begin{figure}
	\includegraphics[width=\columnwidth]{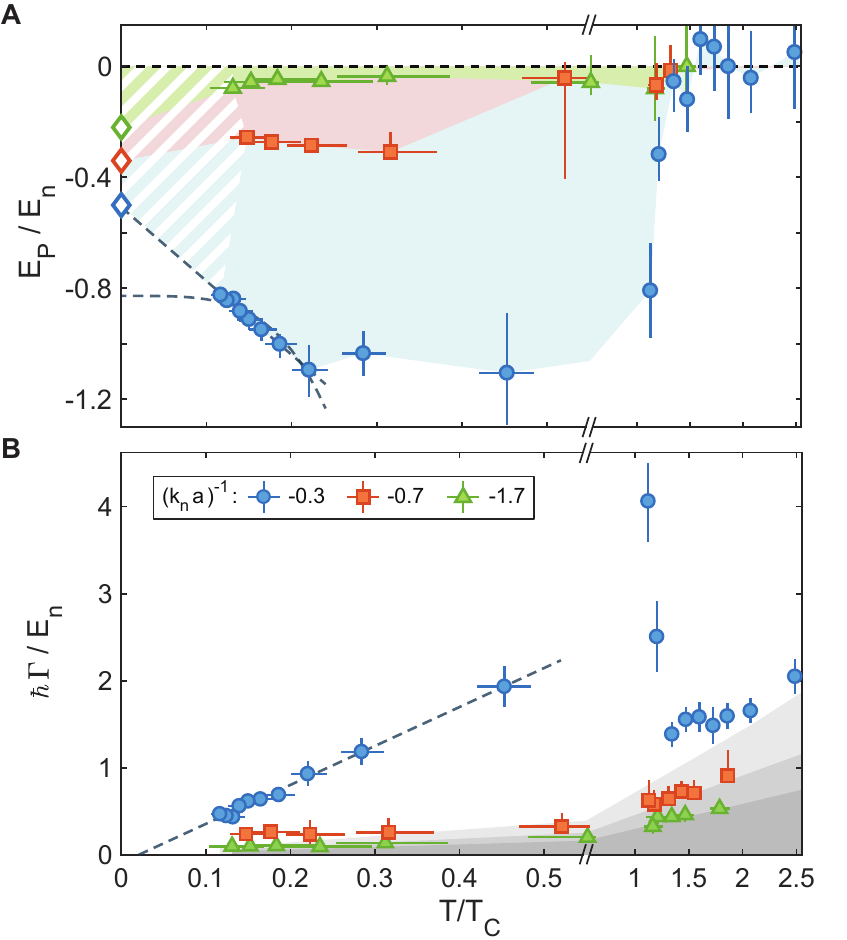}
		\caption{\label{fig:rfvsT}
	Evolution of the Bose polaron as a function of the local reduced temperature $T/T_\mathrm{C}$ for various peak interaction strengths $(k_n a)^{-1}$~\cite{Supplemental}.
	\textbf{(A)} Energy of the Bose polaron. The shaded areas are a guide to the eye and the blue dashed lines represent linear and quartic extrapolations to zero temperature. The prediction of the lowest-order T-matrix calculation is represented by open diamonds at $T\,{=}\,0$.
	\textbf{(B)} The inverse lifetime of the Bose polaron, represented by the half-width at half-maximum ($\Gamma$) of the local rf spectra~\cite{Supplemental}.
	The grey shaded areas indicate the spectral resolutions of the corresponding rf pulses. The dashed line is a linear fit to the data below $T_\mathrm{C}$.}
\end{figure}

Interpreting the spatially-resolved spectrum under the assumption of the local density approximation (LDA)~\cite{Supplemental,Ketterle2008} gives access to the rf spectrum of the impurity as a function of the condensate's local chemical potential $\mu(z) \,{=}\, \mu_0 - V_{\rm Na}(z)$.
Here, $\mu_0 \,{=}\, 4\pi\hbar^2 a_{\rm BB} n_{\rm Na}/m_{\rm Na}$ is the condensate's chemical potential at its peak density, and $V_{\rm Na}(z)$  is the radially-centered trapping potential along the axial direction.
Fig.~\ref{fig:rf}A shows the rf spectrum as a function of $\beta \mu(z)$, the chemical potential normalized by $\beta\,{=}\,1/k_B T$.
The interaction parameter $(k_n a)^{-1}$ also varies with the local density $n_\mathrm{Na}(z)$ as indicated.
A strong shift of the rf transfer for positive chemical potentials is clearly visible.
Fig.~\ref{fig:rf}B shows a selection of spectra, indicating the temperature $T$ normalized by the local critical temperature $T_\mathrm{C}(z) \,{=}\, 3.31 \frac{\hbar^2}{k_B m_\mathrm{Na}}(n_\mathrm{Na}(z))^{2/3}$ for a homogeneous gas.
The absolute frequency of the spectral peak continuously decreases with higher reduced temperatures (left panel). However, when normalized by the degeneracy energy scale $E_n$, the spectral peak frequency in fact increases, indicating a more strongly-bound impurity with increasing temperatures up to the critical temperature $T_\mathrm{C}$ (right panel).
This finding is summarized in Fig.~\ref{fig:rfvsT}A, where the peak frequency shift $\omega_p$ is interpreted as the ground-state energy $E_p\,{=}\,{-} \hbar \omega_p$ of the Bose polaron~\cite{Supplemental}.
Stronger binding of the impurity to the bosonic bath with increasing temperature has been predicted~\cite{Guenther2018}. Additionally, a broadening of the spectral function underlying the rf spectrum may contribute to the observed shift~\cite{Yan2019}.
Above $T_\mathrm{C}$ the peak energy shift suddenly jumps to zero, despite the near-unitarity-limited interactions.
This behavior is expected when the temperature exceeds the energy difference between the attractive and repulsive branches of the resonantly interacting impurity, which occurs near the onset of quantum degeneracy~\cite{Ho2004,Fletcher2017}.
A similar jump in binding energy was recently observed for an impurity resonantly interacting with a nearly degenerate Fermi gas~\cite{Yan2019}.
At weaker attractive interaction, we observe that the Bose polaron is less strongly bound to the bath, as expected~\cite{Rath2013,Li2014} (see Fig.~\ref{fig:rfvsT}A).

In the strongly interacting regime where $(k_n a) \,{\gg}\, 1$, our measurements probe a regime where the binding energy is much larger than the condensate's local mean-field energy.
In this regime, a universal description for the Bose polaron at low temperatures emerges from a lowest order T-matrix and an equivalent variational approach~\cite{Rath2013,Li2014,Supplemental}: here, the impurity acquires an energy shift that is the sum of the individual and uncorrelated shifts from each host boson:
\begin{equation}
E_p \equiv -\frac{\hbar^2 \kappa^2}{2 m_r} = -\frac{2\pi\hbar^2 n_\text{Na}}{m_r} f(i\kappa)
\label{eq:polen}
\end{equation}
where $f(i\kappa) \,{=}\, {-}\frac{a}{1 - \kappa a}$ is the two-body scattering amplitude evaluated at  imaginary momentum $i\kappa$, as appropriate for a bound state.
The equation implicitly gives $E_p$, whose natural energy scale is confirmed as the degeneracy energy scale $E_n$ for an effective particle of reduced mass $m_r$ and density $n_\text{Na}$. In this scenario, $E_p/E_n$ is a universal function of $(k_n a)^{-1}$ only. For weak attractive impurity-boson interactions ($(k_n a)^{-1} {\ll} {-}1$) one finds the mean-field result $E_p \,{=}\, 2\pi\hbar^2 n_\text{Na} a/m_r$, while on the molecular side of the Feshbach resonance in the limit $(k_n a)^{-1} \,{\gg}\, 1$, the polaron energy becomes equal to the energy of a two-body molecule of size $a$, $E_p \,{=}\, {-} \frac{\hbar^2}{2 m_r a^2}$.
On resonance, the approach yields $E_p/E_n \,{=}\, {- }0.71$, which is similar to the result for the unitary Fermi polaron, $E_p/E_n \,{=}\, {-}0.61$~\cite{Chevy2006,Combescot2007}.
The Bose polaron is more strongly bound than its fermionic counterpart due to the lack of constraints imposed by Pauli exclusion~\cite{Supplemental}.
The polaron energies according to the T-matrix approach for $T\,{=}\,0$ are indicated as open diamonds in Fig.~\ref{fig:rfvsT}A.
A linear extrapolation to zero temperature of our strong-coupling binding energy data appears to agree well with this theory.
Alternatively, assuming the increase in binding strength with temperature is due to coupling to the BEC's finite temperature phonon bath, one may attempt a $T^4$ fit to the data~\cite{Levinsen2017}.
Both the linear and quartic extrapolations exclude a simple mean-field prediction that yields $E_p/E_n \,{=}\, {-}1.4$ for $(k_n a)^{-1} \,{=}\, {-}0.3$.

The binding energy alone does not reveal whether the impurities in the bosonic bath form well-defined quasiparticles.
This also requires knowledge of the impurities' spectral width, a measure of the quasiparticle's decay rate~\cite{Schirotzek2009,Yan2019,Zwerger2016,Schneider2009}.
Generally, the width of an rf spectrum corresponds to the rate at which the coherent evolution of an atomic spin is interrupted during the rf pulse.
For quasiparticles, it is momentum-changing collisions with host bosons that cause such decoherence, the same process that limits the quasiparticle's lifetime. The rf spectral width thus directly measures the inverse lifetime of the quasiparticles~\cite{Schirotzek2009,Yan2019,Zwerger2016,Schneider2009}.
Fig.~\ref{fig:rfvsT}B shows that the strong-coupling impurity's spectral width follows a linear dependence with temperature, and strikingly at the Planckian scale: $\Gamma \,{=}\, 8.1(5)\, k_B T/\hbar  $. Observing decay rates at this scale is a hallmark of quantum critical behavior~\cite{Sachdev2011}.
The observed linear trend suggests a well-defined quasiparticle with vanishing spectral width in the limit of zero temperature.
However, near $T_\mathrm{C}$, the rf spectral width increases significantly beyond the measured binding energy $E_p$, signaling a breakdown of the quasiparticle picture.
We attribute both the linear temperature dependence at the Planck scale $k_B T/\hbar$ and the quasiparticle breakdown to the proximity of the Bose-Fermi mixture's quantum multi-critical points~\cite{Ludwig2011,Supplemental}:
the impurity gas is close to the quantum phase transition between the vacuum of impurities, $n_K\,{=}\,0$, and the phase at non-zero impurity density, $n_K\,{>}\,0$; interactions are tuned near the resonant point $(k_n a)^{-1} \,{\rightarrow}\, 0$; and the host boson gas traverses its own quantum critical regime near the onset of quantum degeneracy at $\mu_B \,{\rightarrow}\, 0$.
Here, as only one relevant energy scale remains $(k_B T \,{\approx}\, k_B T_\mathrm{C} \,{\approx}\, 0.55\, E_n$~\cite{Footnote}), the spectral width also scales as $E_n/\hbar$ and no quasiparticles are predicted to persist~\cite{Nikolic2007, Sachdev2011}.
In this regime, the impurities have the shortest mean-free path possible with contact interactions, \textit{i.e.}, one interboson distance.
For all temperatures $T\,{<}\,T_\mathrm{C}$, the scattering rate at the Planckian scale naturally emerges, assuming polarons scatter with thermal excitations of the saturated Bose gas, at density $n_{\rm th} \,{\sim}\, 1/\lambda_B^3$. Given a unitarity-limited scattering cross section $\sigma\,{\sim}\,\lambda_{\rm rel}^2$ and the most probable relative scattering speed $v_{\rm rel} \,{\propto}\, \sqrt{\frac{k_B T}{m_r}}$, we derive a rate
$\Gamma \,{=}\, n_{\rm th}\, \sigma\, v_{\rm rel} \,{\sim}\, \left(m_B/m_r\right)^{3/2} k_B T/\hbar$~\cite{Supplemental}.
Here, $\lambda_{B/\rm rel}$ are the thermal de Broglie wavelengths at the boson and the reduced mass, respectively.
At weaker interaction strengths where $\sigma\,{\sim}\, a^2$, the above relation for $\Gamma$ yields a non-universal rate $\Gamma \,{\propto}\,a^2 T^2$~\cite{Levinsen2017,Supplemental}. Experimentally, the spectral width drops rapidly for the weaker interaction strengths probed here, down to our resolution limit, prohibiting us from discerning the scaling with temperature (see Fig.~\ref{fig:rfvsT}B).

\begin{figure}
	\includegraphics[width=\columnwidth]{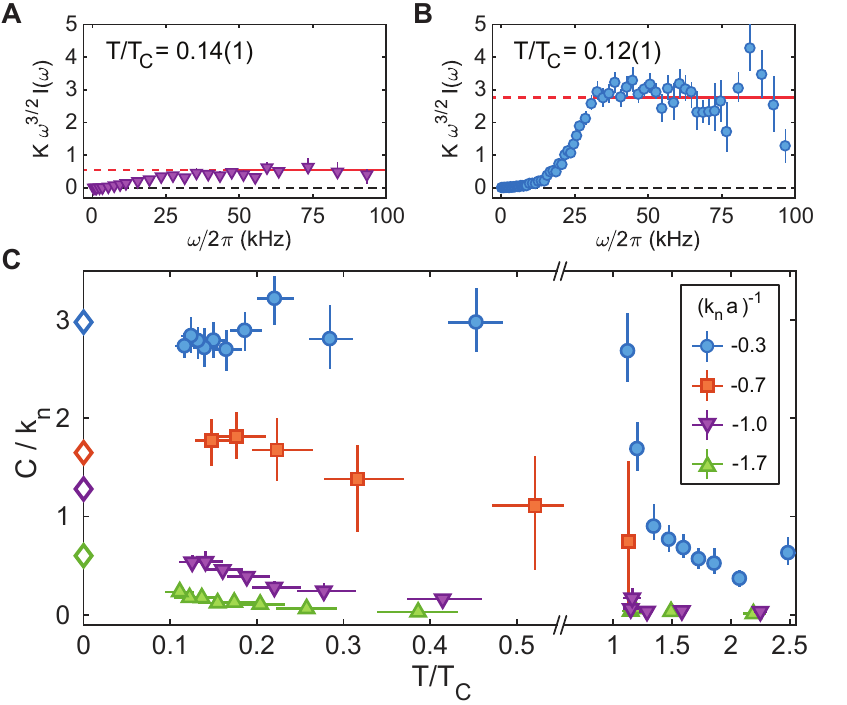}
	\caption{\label{fig:Contact} Contact of the Bose polaron.
	The low temperature rf transfer for \textbf{(A)} $(k_n a)^{-1}\,{=}\,{-}1$ and \textbf{(B)} $(k_n a)^{-1}\,{=}\,{-}0.3$ multiplied by
	$K\omega^{3/2}$, with $K \,{=}\,\frac{8\sqrt{2\pi m_r}}{\Omega_P^2\sigma\sqrt{\hbar}}\frac{1}{k_n}$, displays a plateau that yields the normalized contact $C/k_n$. The contact is obtained from fits in the frequency region indicated by the solid red line.
	\textbf{(C)} The contact, normalized by $k_n$, as a function of the reduced temperature at various interaction strengths. The open diamonds at $T\,{=}\,0$ are the T-matrix predictions from Eq.~\ref{eq:contact}.}
	\vspace{-1em}
\end{figure}

Spectra obtained via rf {ejection} spectroscopy encode the wavefunction overlap between the interacting, dressed impurity and a non-interacting state. As such, they not only contain information about the binding energy and lifetime of the impurity, but also about the short-range correlations between the impurity and the surrounding medium.
Indeed, the final state of an {ejection} spectrum at high rf frequencies is a free impurity with large momentum $\hbar k$.
Hence, high frequencies in {ejection} spectroscopy probe the initial wavefunction at short distances~\cite{Baym2007,Punk2007,Braaten2010,Zwierlein2016}.
This leads to characteristic tails ${\propto}\, \omega^{-3/2}$ of the rf spectra, reflecting the two-body nature of the wavefunction at short distances. The strength of the rf transfer is directly proportional to the contact $C$, a thermodynamic quantity of the many-body system describing the probability that the impurity is in close vicinity to the host bosons. Through the Hellmann-Feynman theorem, the contact is also equal to the change in energy with the inverse impurity-boson scattering length: $\frac{C}{k_n} \,{=}\, 2\pi \frac{d}{d(k_na)^{-1}}(\frac{{-}E_p}{E_n})$~\cite{Baym2007,Punk2007,Tan2008,Zwierlein2016}.
The high-frequency tail of the rf transfer $I(\omega)$ can be written~\cite{Braaten2010}:
\begin{equation}
\label{eq:contactFromRF}
I(\omega)\underset{\omega\rightarrow\infty}{=}\frac{\Omega_P^2\sigma}{8}\sqrt{\frac{\hbar}{2\pi m_\text{r}}}\frac{C}{\omega^{3/2}}
\end{equation}
where $\sigma$ is the Gaussian $e^{-1/2}$ width of the rf pulse's duration and $\Omega_P$ the peak Rabi frequency~\cite{Supplemental}.
Our spectra follow this behavior closely: when multiplied by $\omega^{3/2}$, they asymptote to plateaus that yield the contact's value, shown for $(k_n a)^{-1}\,{=}\,{-}1$ in Fig.~\ref{fig:Contact}A and $(k_n a)^{-1}\,{=}\,{-}0.3$ in Fig.~\ref{fig:Contact}B, respectively.
Fig.~\ref{fig:Contact}C summarizes our measurements of the contact, normalized by $k_n$, as a function of $T/T_\mathrm{C}$ for various interaction strengths.
From weak to strong attractive interactions, the contact increases monotonically.
For the strongest interaction strength the normalized contact remains approximately constant for all values $T/T_\mathrm{C} \,{<}\, 1$. An abrupt drop of the normalized contact is seen above the BEC transition temperature, though it remains non-zero.
This is expected for a Boltzmann gas with unitarity-limited interactions that has a non-zero contact given by the inverse mean-free path, $n\sigma \,{\propto}\, 1/T$.
Therefore $C/k_n$ decreases as $T_\mathrm{C}/T$ in the non-degenerate regime~\cite{Fletcher2017, Yan2019}.
The low-temperature value of the normalized contact is close to what one finds for the unitary Fermi polaron ($C/k_n \,{=}\, 4.3$~\cite{Yan2019}), the balanced unitary Fermi gas~\cite{Lingham2016,Mukherjee2019,Carcy2019}, and the near-unitary BEC ~\cite{Wild2012}.
Using the variational ansatz for the Bose polaron's energy (see Eq.~\ref{eq:polen}), we obtain an expression for the normalized contact:
\begin{equation}
\frac{C}{k_n} = \pi^2 \frac{E_p/E_n}{E_n/E_p - \frac{\pi}{4}\frac{1}{k_n a}}
\label{eq:contact}
\end{equation}
which yields $\frac{C}{k_n} \,{=}\, \pi^2 \left(\frac{E_p}{E_n}\right)^2 \,{=}\, 5.0$ on resonance.
The contact can also be interpreted in an intuitive picture~\cite{Tan2008}: it gives the number of bosons $N_B(s)$ within a sphere of radius $s$ around the impurity: $N_B(s) \,{=}\, C s/4\pi$ for $s\,{\ll}\, n_{\rm{Na}}^{-1/3}$ and $s\,{\ll}\, |a|$. The measured near-unity value of $C/4\pi$ in units of the interboson spacing thus indicates that even for near-unitarity-limited interactions -- on average -- only about one extra boson is in close proximity to the impurity.  In this respect, the resonant Bose polaron shares traits with a molecular dimer of a size given by the interboson distance. Within the variational description, the localized part of the polaron's wavefunction is of identical form to that of a molecule, and away from resonance where $a\,{>}\,0$, the polaron smoothly evolves into a molecule of size $a$~\cite{Rath2013,Li2014, Supplemental}.

For future studies, it will be interesting to probe transport properties of the impurities and specifically investigate whether their resistivity scales linearly with temperature, in analogy to findings in the strange metal phase of the cuprates~\cite{Lee2006}.
Furthermore, increasing the impurity concentration might allow the formation of bipolarons~\cite{Camacho-Guardian2018} and the observation of phonon-induced superfluidity~\cite{Efremov2002, Kinnunen2018}.

The authors thank Richard Schmidt, Fabian Grusdt, Yulia Shchadilova, Kushal Seetharam, Eugene Demler, Eberhard Tiemann, and Wilhelm Zwerger for useful discussions. We also thank Joris Verstraten, Alexander Chuang, and Elisa Soave for experimental assistance, and Biswaroop Mukherjee, Zhenjie Yan, and Richard Fletcher for critical reading of the manuscript.  This work was supported by the NSF, AFOSR, ARO, an AFOSR MURI on ``Exotic Phases of Matter'', the David and Lucile Packard Foundation, and the Gordon and Betty Moore Foundation through grant GBMF5279.  Z. Z. Y. acknowledges support from the NSF GRFP.

\bibliography{References}% Produces the bibliography via BibTeX.

\begin{thebibliography}{84}
\bibitem [{Foo()}]{Footnote}%
  \BibitemOpen
  \href@noop {} {}\bibinfo {note} {Efimov trimers are not expected to enter in
  the description of our system at quantum criticality, as their size is
  predicted to be orders of magnitude larger than the interparticle spacing at
  our densities (see [40]).}\BibitemShut {Stop}%
\end{thebibliography}
%%For suppmat
\renewcommand{\thefigure}{S\arabic{figure}}
\setcounter{figure}{0}
\setcounter{equation}{0}

\clearpage
\onecolumngrid
\vspace{\columnsep}

	\begin{center}
	\large{\textbf{Supplementary material: Bose polarons near quantum criticality}}\\~\\

	Zoe Z. Yan$^{1}$, Yiqi Ni$^{1}$, Carsten Robens$^{1}$, and Martin W. Zwierlein$^{1}$\\~\\

	\textit{$^{1}$MIT-Harvard Center for Ultracold Atoms, Research Laboratory of Electronics, and Department of Physics,
	Massachusetts Institute of Technology, Cambridge, Massachusetts 02139, USA}

	\end{center}

\vspace{\columnsep}
\twocolumngrid

\begin{center}
	\textbf{Experimental methods}
\end{center}
\smallskip
The preparation of the ultracold mixture of $^{23}$Na and $^{40}$K closely follows the process described in Ref.~\cite{Park2012}.
In brief, the $^{40}$K impurities and the bosonic $^{23}$Na are cooled and trapped in a crossed optical dipole trap with wavelength $\lambda\,{=}\,\SI{1064}{\nano\meter}$.  The Na trap frequencies are $(110, 78, 13)$ \SI{}{\hertz} in the x-,y-, and z-directions.  The atoms are prepared in $\ket{F,m_F}=\ket{1,1}$ and $\ket{9/2,-9/2}$ for $^{23}$Na and $^{40}$K, respectively. $F$ is the total angular quantum number and $m_F$ is its projection along the magnetic field axis. After evaporative cooling in the optical trap, the $^{23}$Na cloud of $\approx 10^6$ atoms has undergone Bose-Einstein condensation and sympathetically cooled the impurities to a final temperature of \SI{130}{\nano\kelvin}.
We adiabatically ramp on an additional single-beam optical dipole trap at $\lambda\,{=}\,\SI{775}{\nano\meter}$ along the laboratory $z-$axis (the axial coordinate described in the main text), which provides additional confinement for the impurities but not the bosonic atoms due to their differing ac polarizabilities.  This species-selective optical dipole trap cancels the differential gravitational sag between the two species.
The impurity atoms form a cloud with Gaussian widths of (9,9,70) \SI{}{\micro\meter} in the x-,y-, and z-axes, respectively.
\begin{figure}[tb]

		\includegraphics[width=\columnwidth]{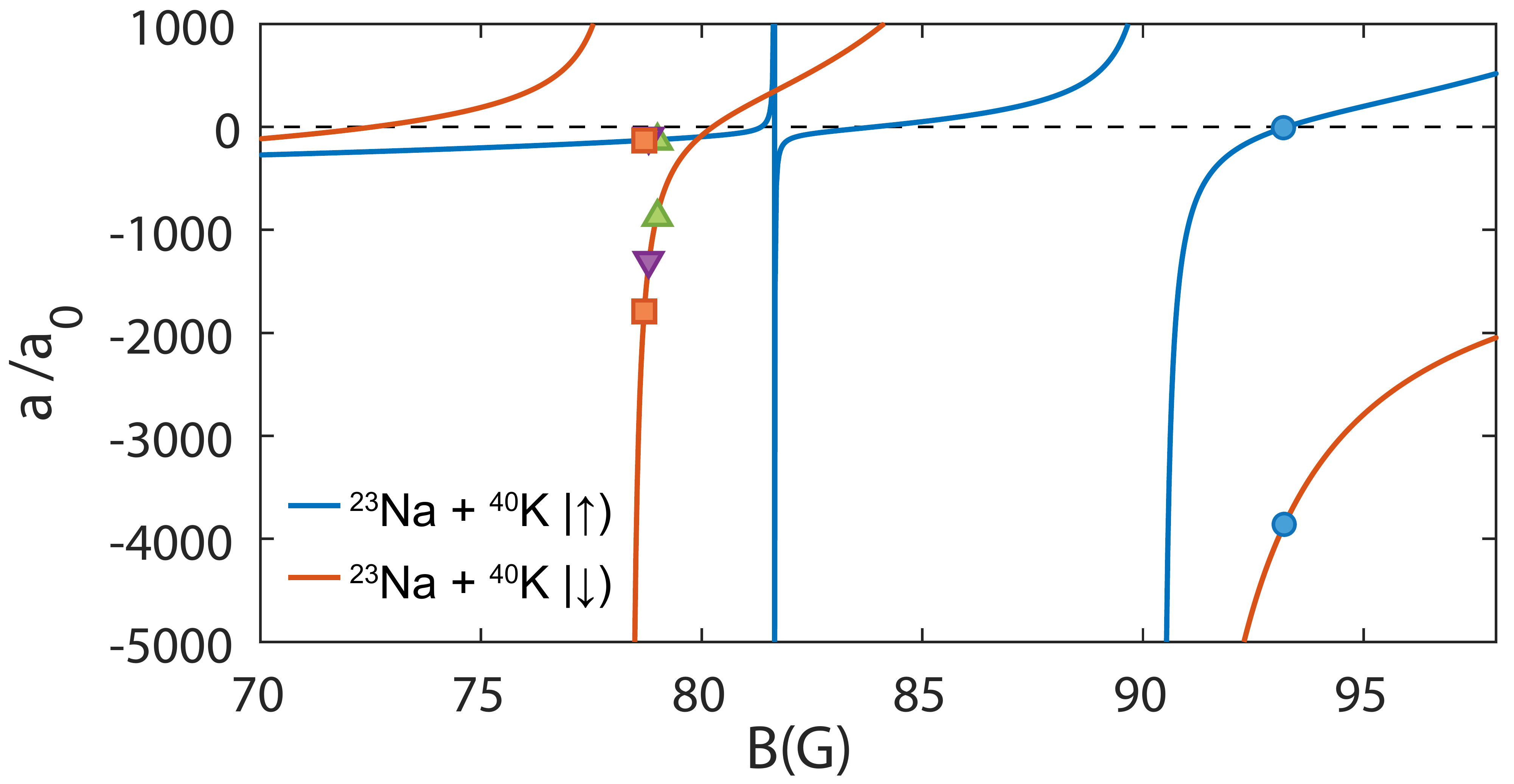}
		
		\caption{\label{fig:FB} The impurity-bose s-wave scattering lengths versus magnetic field for the two impurity spin states $\ket{\uparrow},\ket{\downarrow}$.  Markers denote the magnetic fields where we perform rf spectrsocopy for various $a$, taking care to keep the magnitude of $a_\uparrow$ small. At \SI{93.2}{\gauss} we exploit the zero-crossing of the $\ket{\uparrow}$ state to avoid any final state interactions.
		}

\end{figure}
To tune interactions between the impurities and the bath, we make use of the Feshbach resonances between $^{23}$Na and $^{40}$K, shown in Fig.~\ref{fig:FB}.  Our previous measurement of the interspecies resonances using Feshbach loss spectroscopy was reported in Ref.~\cite{Park2012}. For this study we refined the location of the resonances and zero-crossing of the scattering length using a technique based on interspecies thermalization (similar to the method in Ref.~\cite{Ohara2002}).

\begin{figure}[tb]

		\includegraphics[width=\columnwidth]{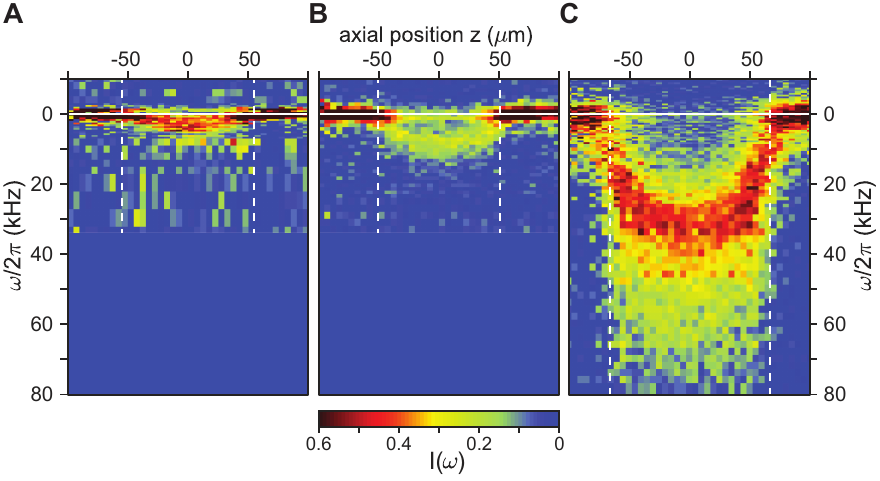}
		
		\caption{\label{fig:sidebyside} Color density maps of the impurity transfer $I(\omega)$ versus rf frequency and cloud axial location, with the boson Thomas-Fermi radii denoted by the white dashed lines, for scattering lengths of \textbf{(A)} $-840\,a_0$, \textbf{(B)} $-1800\,a_0$, and \textbf{(C)} $-3900\,a_0$.  These scattering lengths correspond to peak interaction strengths $(k_\mathrm{n} a)^{-1}{ = }$ -1.7, -0.7, and -0.3, respectively, at the cloud center.  In all cases, the rf drive was a Gaussian pulse lasting \SI{500}{\micro\second}, with the Gaussian $\sigma = \SI{62.5}{\micro\second}$.  The peak Rabi frequencies on the bare atomic lines were $2\pi\times$\SI{3.5}{\kilo\hertz} for (a)-(b) and $2\pi\times$\SI{11}{\kilo\hertz} for (c).
		}

\end{figure}

In preparation for spectroscopy, we ramp the magnetic field to the vicinity of a Feshbach resonance, where the impurity-bath interaction can be tuned from weakly- to strongly-attractive.
For the measurements performed with near-unitarity-limited interactions at $a\,{=}\,{-}3900\,a_0$ (cf. Figs.~$1\,{-}\,4$ in the main text), the final magnetic field is \SI{93.2}{\gauss}, where the final state $\ket{\uparrow}$ has zero interaction with the bosons.  For the data at scattering lengths $a\,{=}\,{-}840,{-}1800\,a_0$ in the initial state (cf. Figs.~$3\,{-}\,4$ in the main text), the field is varied around a Feshbach resonance near \SI{78.4}{\gauss}, where the final state is weakly interacting ($a_\uparrow\,{\approx}\,{-}100\,a_0$).
For ejection spectroscopy, we use an rf pulse to transfer the impurities out of the interacting state $\ket{\downarrow}$ into a non-interacting state $\ket{\uparrow}$.  The Gaussian rf pulse is symmetrically truncated after a duration of $8\sigma$, where $\sigma\,{=}\,$\SI{62.5}{\micro\second} is the temporal $e^{-1/2}$ width of the Gaussian pulse.
The spectral resolution of our pulse is dominated by Fourier broadening, leading to a pulse full-width-half-maximum (FWHM) of $\approx\,$\SI{6}{\kilo\hertz}.
After the rf probe, we immediately image the two impurity states and the bosons, employing \textit{in situ} absorption imaging.  Sample spectra for three different interaction strengths are shown in Fig.~\ref{fig:sidebyside}.  As described in the main text, the stronger impurity-bath couplings correspond to a greater shift in $E_p$.

For error estimation in Fig.~2, the errorbars reflect standard errors for the impurity transfer, where for each frequency we took 11 repeated samples.  For the polaron energy in Fig.~3, each point is determined by bootstrapping the rf lineshape 1000 times, then applying a peak-finder algorithm that can take into account the asymmetry of the lineshape.  Error bars denote the 68\% confidence interval of the bootstrapped peak position.  Error bars in the spectral width and contact are determined similarly.  Horizontal errorbars reflect the statistical uncertainty both in absolute temperature $T$ and in local critical temperature $T_\mathrm{C} \,{\sim}\, n_{\rm Na}^{2/3}$.

\begin{center}
	\textbf{Comparison between ejection and injection rf spectroscopy}
\end{center}
\smallskip
There are two commonly used techniques to gain information on a cold-atom system using rf spectroscopy.
The many-body system can be prepared by initializing the impurities in the non-interacting state and injecting them into the interacting state, a technique known as \textit{injection} or indirect rf spectroscopy.  This method has been employed~\cite{Hu2016,Jorgensen2016} to measure the continuum of excited polaron states.  Alternatively, we can prepare the many-body system in the interacting state in thermal equilibrium and then eject the impurities out of this state (\textit{ejection} or direct rf spectroscopy).  As we will demonstrate in the following, the two techniques lead to different spectral responses: the former method excites the impurity into a continuum of states while the latter probes the impurity's ground state.

\begin{figure}

		\includegraphics[width=\columnwidth]{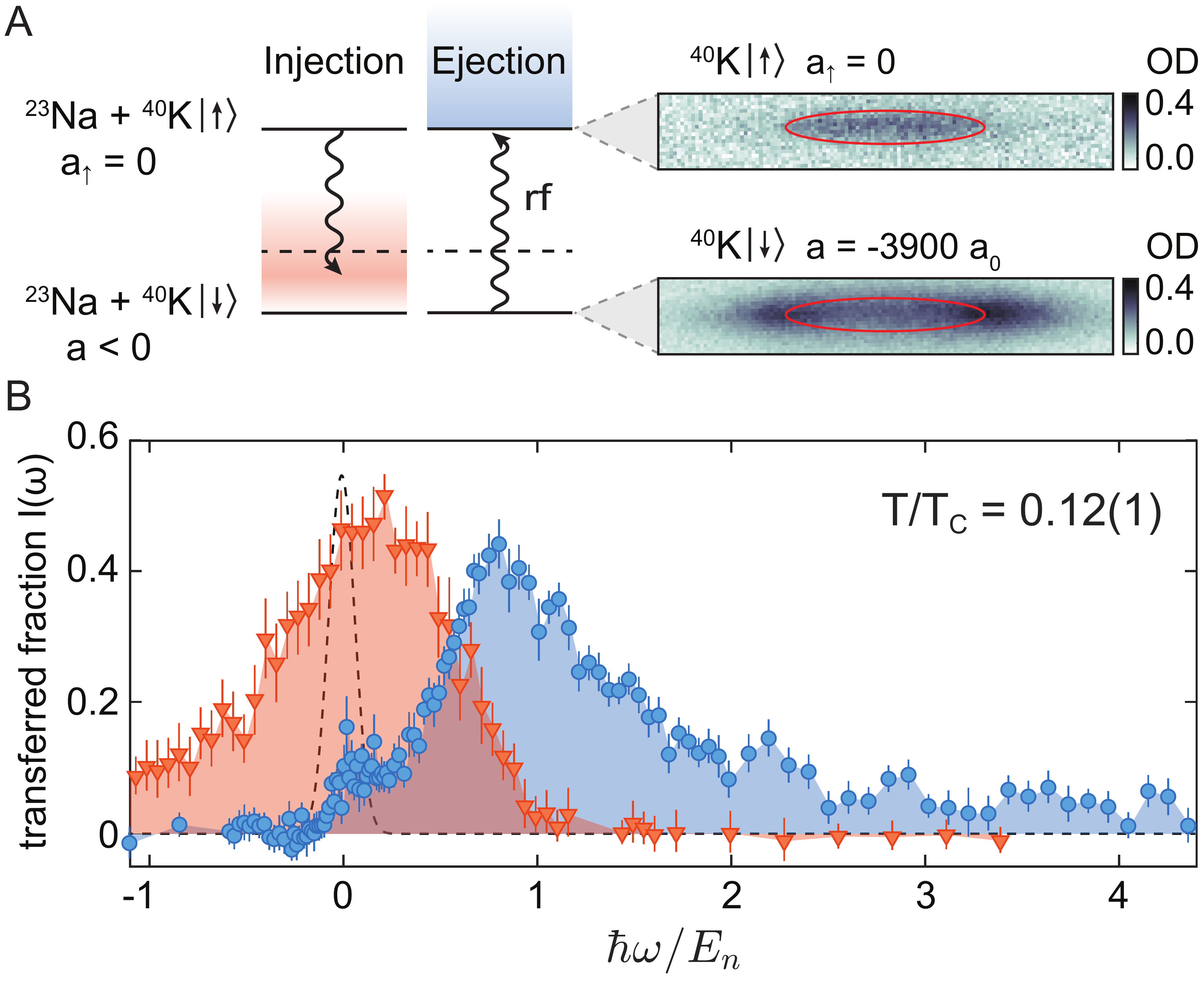}
		\caption{\label{fig:InjVsEj} Spectral response of $^{40}$K impurities using ejection and injection spectroscopy. \textbf{(A)} The energy levels of the many-body system (left).  The dashed lines denote the impurity's hyperfine energies and the solid lines denote the ground state polaron energies.  In the ejection technique, the polaron ground state is prepared and ejected out (pictured on the bottom right as an absorption image of K in $\ket{\downarrow}$) via an rf drive to the non-interacting state $\ket{\uparrow}$ (pictured on the top right.) The BEC's Thomas-Fermi radius is marked by the red line. \textbf{(B)} A comparison of the injection and ejection lineshapes at $a{=} -3900 $ $a_0$, in red and blue respectively, with peak interaction strength $(k_\mathrm{n} a)^{-1} {=} -0.3$.  The transfer from the initial to the final spin state is shown as a function of normalized rf frequency.  The dotted black line shows the Gaussian response of the bare impurity state.
		}

\end{figure}

We compare the two methods at an interaction strength of $-3900\,a_0$, as shown in Fig.~\ref{fig:InjVsEj}.  In the ejection method we employ for the data shown in the main text, the polaron is prepared in equilibrium and ejected into a non-interacting spin state, denoted $\ket{\uparrow}$.  The injection method is also performed with all other parameters (\textit{i.e.}~density, rf pulse profile) held constant.
As shown in Fig.~\ref{fig:InjVsEj}(B), in the injection protocol, the maximum transfer of population occurs at a normalized energy of $0.2\,E_n$, much lower compared to the measured energy shift of $0.8\,E_n$ obtained from ejection spectroscopy.
Thus, assigning the spectral peak of $0.2\,E_n$ as the polaron binding energy would be a significant under-prediction.
It is the onset and not the peak of the injection spectrum that encodes any meaning for the ground state polaron energy $E_p$~\cite{Yan2018Talk,Ardila2019}.
Moreover, only the ejection method can recover additional equilibrium quantities of the Bose polaron, including its lifetime and short-range correlations, as the injection spectrum convolves the spectral response of a continuum of excited polaron states~\cite{Ardila2019}.

\begin{center}
	\textbf{Local boson density and reduced temperature}
\end{center}
\smallskip
The \textit{in-situ} local boson density $n_{\mathrm{Na}}(\mathbf{r})~$-- the sum of the condensate density $n_{c}(\mathbf{r})$ and the thermal density $n_t(\mathbf{r})$ -- has the following form under the local density approximation (LDA) and the Thomas-Fermi limit~\cite{Naraschewski1998}:

\begin{align}
n_{c}(\textbf{r})&= \frac{15 N_c}{8\pi R_xR_yR_z}\max\biggr(1-\frac{x^2}{R_x^2}-\frac{y^2}{R_y^2}-\frac{z^2}{R_z^2},0\biggr )\\
n_t(\textbf{r})&=\frac{1}{\lambda_\mathrm{dB}^3}g_{3/2}\bigg(\exp\big(-\beta \big|\mu_0-\frac{1}{2} m_\mathrm{Na}\sum_{i=x,y,z} \omega_i^2 r_i^2\big|\big)\bigg)
\end{align}

where $\lambda_\mathrm{dB}=\sqrt{2\pi \hbar^2/m_\mathrm{Na}k_BT}$ is the thermal de Broglie wavelength, $g$ the Polylogarithm function, $\beta\equiv 1/k_B T$, and $\mu_0$ the peak boson chemical potential.
$R_i$ is the Thomas-Fermi radius along the $i$th coordinate, defined by $\mu_0=m\omega_i^2R_i^2/2$.
The condensate number is $N_c=\frac{8\pi}{15g_\mathrm{BB}}\mu_0R_xR_yR_z$, where $g_\mathrm{BB}=4\pi\hbar^2a_\mathrm{BB}/m_\mathrm{Na}$ is the bose-bose coupling constant.
The condensate is assumed to only experience mean-field repulsion, while the thermal atoms are an ideal gas confined in the external harmonic trap and the mean-field repulsion of the condensate~\cite{Naraschewski1998}.  Experimental values of $\mu_0$ are derived from the speed of the BEC's hydrodynamic expansion after releasing the cloud in time-of-flight~\cite{Dalfovo1998}.  Temperature is also measured in time-of-flight by fitting the outer wings to the thermal density distribution.
From the local density, we compute the local critical temperature $T_\mathrm{C}$, where $k_BT_\mathrm{C}(\mathbf{r})=\frac{3.31\hbar^2}{m_\mathrm{Na}}n_\mathrm{Na}(\mathbf{r})^{2/3}$.

\begin{center}
	\textbf{Bose Polaron properties at $T=0$ within the variational approach}
\end{center}
\smallskip
Here we obtain the Bose polaron energy and contact from a simple variational ansatz, introduced originally for the description of the Fermi polaron~\cite{Chevy2006}. We work in the limit of weak boson-boson interaction, justified \textit{a posteriori} as we find that the boson chemical potential is much smaller than the polaron energy $E_p$. In this limit, the linear portion of the Bogoliubov spectrum for the bosons at low momenta is not relevant for the energetics of the polaron, and we can simply work with a free-particle dispersion. Our solution turns out to be identical to what is obtained from the lowest-order T-matrix calculation in the same limit~\cite{Rath2013}.

The Hamiltonian describing the impurity interacting with the Bose gas is
\begin{equation}
H = \sum_\vect{k} \left( \eB{k} a^\dagger_\vect{k} a_\vect{k} + \eI{k} c^\dagger_\vect{k} c_\vect{k}\right) + \frac{g_0}{V} \sum_{\vect{k}\vect{k'}\vect{q}} a^\dagger_{\vect{k}+\vect{q}} c^\dagger_{\vect{k'}-\vect{q}} c_{\vect{k'}} a_{\vect{k}}.
\end{equation}
Here, $\eB{k} = \hbar^2 k^2/2 m_B$ and $\eI{k}= \hbar^2 k^2/2 m_I$ are the boson and impurity free dispersions, $m_B$ and $m_I$ the boson and impurity mass, respectively, $a^\dagger_\vect{k}$ and $c^\dagger_\vect{k}$ are the boson and impurity creation operators, and $V$ the quantization volume. $g_0$ is the bare impurity-boson coupling constant, related to the impurity-boson scattering length $a$ via the Lippmann-Schwinger equation
\begin{equation}
\frac{1}{g_0} = \frac{m_r}{2\pi\hbar^2 a} - \frac{1}{V} \sum_\vect{k} \frac{1}{\eB{k}+\eI{k}}.
\end{equation}

\begin{figure}

		\includegraphics[width=\columnwidth]{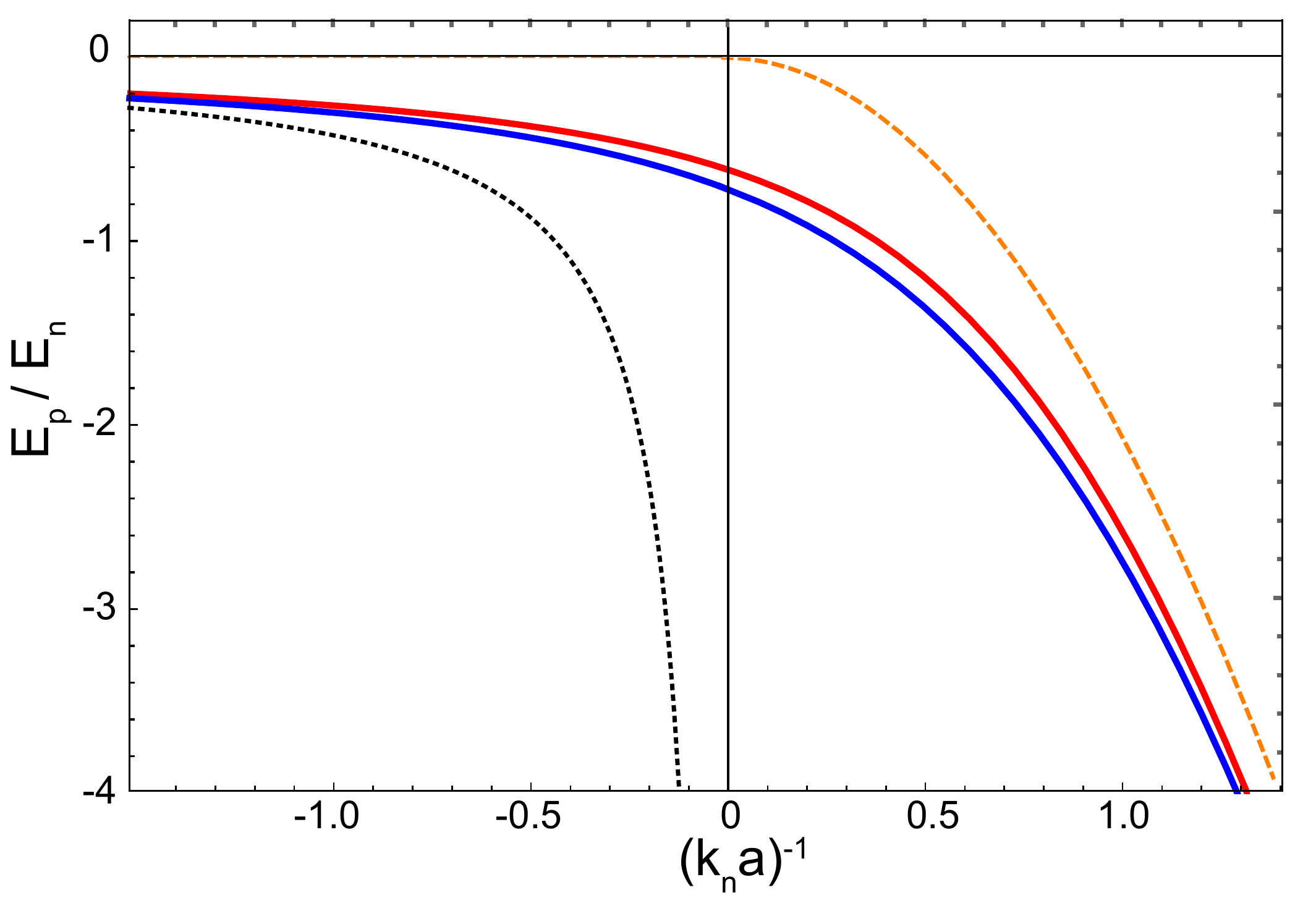}
		
		\caption{\label{fig:variationalEnergy} Binding energy of the Bose and Fermi polaron from the variational ansatz (solid blue and red line, respectively). For comparison, the bare molecular energy is shown (dashed, yellow) as well as the mean-field result (black, dotted).
		}

\end{figure}

The variational wavefunction is a superposition of the unscattered impurity at rest $\left|\vect{0}\right>_I$ immersed in the Bose condensate and the impurity, scattered into momentum state $-\vect{k}$, having ejected one boson out of the condensate into momentum $\vect{k}$:
\begin{equation}
\left|\Psi\right> = \phi_0 \left|\vect{0}\right>_{\rm I} \otimes \left|\alpha\right>_{\rm B} + \sqrt{\NB}\sum_\vect{k} \phi_\vect{k} a^\dagger_\vect{k} \left|-\vect{k}\right>_{\rm I} \otimes \left|\alpha\right>_{\rm B} .
\end{equation}
The factor $\sqrt{\NB}$, where $\NB$ is the average boson number, originates from the destruction operator $a_{\vect{0}}$ acting on the condensate, taken to be in a coherent state $\left|\alpha\right>_{\rm B}$ with $\alpha \,{=}\, \sqrt{\NB}$.
The expectation values of the kinetic and potential energies are
\begin{equation}
\left<H_0\right> = \NB \sum_\vect{k} \left|\phi_\vect{k}\right|^2 (\eB{k}+\eI{k})
\end{equation}
\begin{equation}
\left<V\right> = g_0\, n_B \left(\left|\phi_0\right|^2 + \sum_\vect{k} \left(\phi_0 \phi^*_\vect{k} + \phi^*_0 \phi_\vect{k}\right) + \sum_{\vect{k}\vect{k'}} \phi_\vect{k} \phi^*_\vect{k'}\right)
\end{equation}
with $\nB=\NB/V$ the boson density.
Minimization of the total energy, under the normalization constraint $\left<\Psi|\Psi\right> = 1$, yields the following set of equations:
\begin{eqnarray}
g_0 \nB \chi &=& E_p\phi_0 \\
g_0 \frac{1}{V} \chi &=& (E_p - \eB{k} - \eI{k})\phi_\vect{k}
\end{eqnarray}
with $\chi = \phi_0 + \sum_\vect{k} \phi_\vect{k}$ and $E_p$ the polaron energy.
One thus obtains for the coefficients $\phi_\vect{k}$:
\begin{equation}
\phi_\vect{k} = \frac{1}{\NB} \frac{E_p}{E_p-\eB{k}-\eI{k}}\phi_0,
\end{equation}
and for the polaron energy $E_p$, after eliminating $g_0$ in favor of the scattering length $a$:
\begin{equation}
E_p = \frac{\nB}{\frac{m_r}{2\pi\hbar^2 a} - \frac{1}{V}\sum_\vect{k}\left(\frac{1}{E_p - \eB{k}-\eI{k}} + \frac{1}{\eB{k}+\eI{k}}\right)}
\end{equation}
Noting that $\frac{1}{V}\sum_\vect{k}\left(\frac{1}{E_p - \eB{k}-\eI{k}} + \frac{1}{\eB{k}+\eI{k}}\right) = \frac{m_r \kappa}{2\pi \hbar^2}$, where we have set $E_p \equiv - \hbar^2 \kappa^2/2 m_r$, one obtains the Eq.~1 of the main text:
\begin{equation}
E_p \equiv -\frac{\hbar^2\kappa^2}{2 m_r} = -\frac{2\pi\hbar^2 \nB}{m_r} f(i\kappa)
\end{equation}
where $f(i\kappa)=-\frac{a}{1-\kappa a}$ is the two-body scattering amplitude at imaginary momentum $i\kappa$.
Fig.~\ref{fig:variationalEnergy} compares the variational polaron energy of the Bose polaron with that of the Fermi polaron -- an impurity immersed in a Fermi sea of the same density. The difference in energies is generally less than $\sim 0.2 E_n$ for all interaction strengths. The Fermi polaron is less strongly bound due to the spread of relative momenta in the initial state and Pauli blocking of final scattering states.

\begin{figure*}
		\includegraphics[width=2\columnwidth]{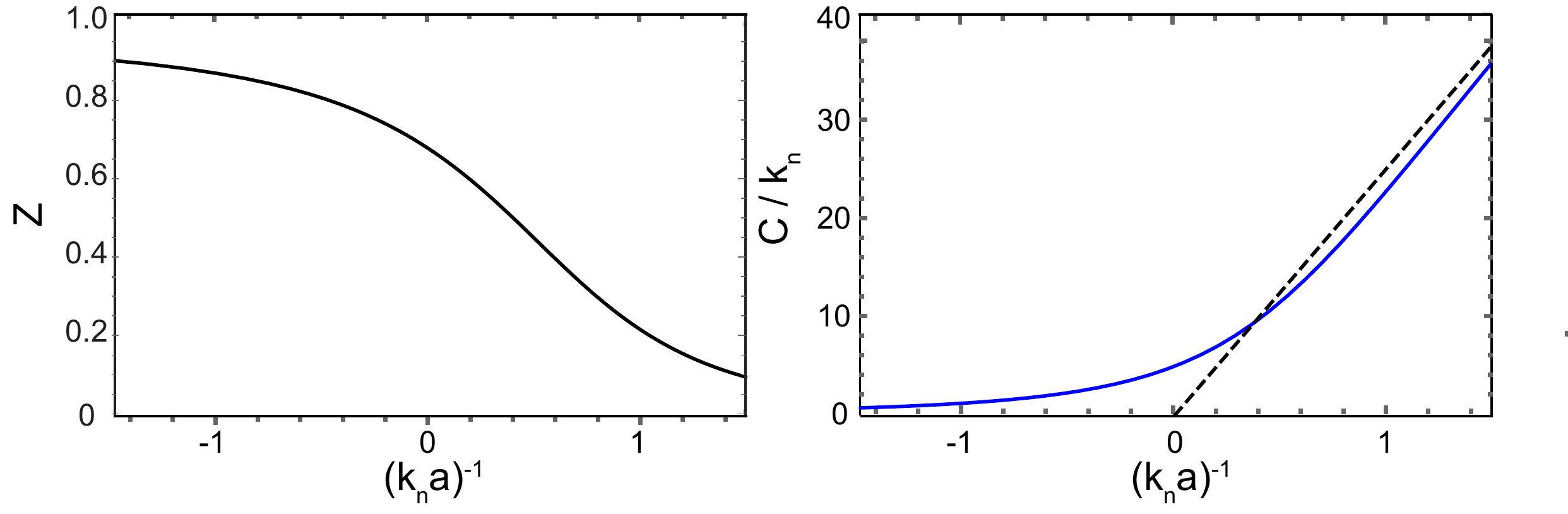}
		
		\caption{\label{fig:contactBP} The residue (left) and contact (right, blue curve) of the Bose polaron, within the variational ansatz. The contact for a bare molecule is shown as the dashed black line.
		}
\end{figure*}

The normalization condition $\left|\phi_0\right|^2 + \NB \sum_\vect{k} \left|\phi_\vect{k}\right|^2 = 1$ yields the quasiparticle weight in this approximation:
\begin{equation}
Z \equiv \left|\phi_0\right|^2 = \frac{1}{1+\frac{1}{8\pi} \frac{\kappa^3}{\nB}} = \frac{1}{1+\frac{1}{2}\left(1-\frac{E_p}{E_{\rm mf}}\right)}
\end{equation}
$Z$ is thus simply related to the ratio of $E_p$ and the mean-field result $E_{\rm mf} \equiv \frac{2\pi\hbar^2 \nB a}{m_r}$.
On resonance, where $1/E_{\rm mf} = 0$, this approach yields a quasiparticle weight of $Z = 2/3$, as seen in Fig.~\ref{fig:contactBP}.

The contact $C \equiv \frac{8\pi m_r}{\hbar^2} \frac{\partial E_p}{\partial\left(-a^{-1}\right)}$ directly follows from the $a$-dependence of $E_p$ as
\begin{equation}
\frac{C}{k_n} = \pi^2 \frac{\frac{E_p}{E_n}}{\frac{E_n}{E_p}-\frac{\pi}{4}\frac{1}{k_n a}}
\end{equation}
The momentum distribution of the polaron is given by a delta-function centered at $\vect{k}=0$, of weight $Z$, plus a contribution $n_k = \NB \left|\phi_k\right|^2$ from impurity-boson scattering:
\begin{eqnarray}
n_k &=& \frac{Z}{\NB} \frac{E_p^2}{\left(E_p - \eB{k}-\eI{k}\right)^2} \nonumber \\
&=& \frac{Z}{\NB} \frac{1}{\left(1+ \frac{k^2}{\kappa^2}\right)^2}
\end{eqnarray}
This component of the momentum distribution has the same dependence on momentum as that of a Feshbach molecule of spatial size $1/\kappa$, the length scale set by the polaron energy.
In real space, the wavefunction of the polaron thus is a superposition of a delocalized wave of weight $Z$ and a part that is localized, of the form $\frac{1}{r} \exp(-\kappa r)$.

The contact $C$ governs the high-momentum tails of the momentum distribution according to $n_k \rightarrow \frac{C/V}{k^4}$. From that we obtain a relation between the contact, the quasiparticle weight and the polaron energy (see Fig.~\ref{fig:contactBP}), valid within this variational approach:
\begin{equation}
\frac{C}{k_n} = \frac{3\pi^2}{2} Z \left(\frac{E_p}{E_n}\right)^2
\end{equation}

\paragraph{\bf{Rf spectrum of the Bose polaron:}}
Fermi's Golden Rule yields for the radiofrequency spectrum of the Bose polaron:
\begin{equation}
\Gamma_{\rm rf}(\omega) = \frac{2\pi}{\hbar} \sum_f \left|\left<f\right|V_{\rm rf}\left|\Psi\right>\right|^2 \delta\left(\hbar\omega - (E_f - E_p)\right)
\end{equation}
where $V_{\rm rf} = \frac{1}{2}\hbar\Omega_R \sum_\vect{k} d^\dagger_\vect{k} c_\vect{k} + {\rm c.c.}$, $\Omega_R$ is the Rabi frequency, and the sum extends over a complete set of final states $\left|f\right>$ of energy $E_f$. With $d^\dagger_\vect{k}$ the creation operator for an impurity atom in the final, non-interacting state of the rf transition, the final states are
\begin{eqnarray}
\left|\vect{0}\right> &\equiv& d^\dagger_0 \left|{\rm vac}\right>_{\rm I} \otimes \left|\alpha\right>_{\rm B} \nonumber \\
\left|\vect{k}\right> &\equiv& d^\dagger_{-\vect{k}} a^\dagger_\vect{k} \left|{\rm vac}\right>_{\rm I} \otimes \left|\alpha\right>_{\rm B}.
\end{eqnarray}
where $\left|{\rm vac}\right>_{\rm I}$ is the vacuum of the impurity states. The energies of the states $\left|\vect{k}\right>$ are $E_f = \eB{k}+\eI{k}$. For the coupling matrix elements one has
\begin{eqnarray}
\left<\vect{0}\left|V_{\rm rf}\right|\Psi\right> &=& \frac{\hbar\Omega_R}{2} \phi_0 \nonumber \\
\left<\vect{k}\left|V_{\rm rf}\right|\Psi\right> &=&
\frac{\hbar\Omega_R}{2} \phi_k \alpha
\end{eqnarray}
One then finds for the normalized spectrum $\tilde{I}(\omega)\equiv \frac{\Gamma_{\rm rf}(\omega) E_n}{\frac{\pi}{2} \hbar \Omega_R^2}$:
\begin{equation}
\label{eq:tmatrix}
\tilde{I}(\omega) = Z \delta\left(\frac{\hbar \omega}{E_n} {+}\frac{E_p}{E_n}\right)+ \frac{1}{2\pi^2\sqrt{2}} \frac{C}{k_n} \left(\frac{E_n}{\hbar\omega}\right)^2 \sqrt{\frac{\hbar \omega}{E_n} {+}\frac{E_p}{E_n}}
\end{equation}

The spectrum has a delta-component shifted by the polaron energy, and a background from impurity-boson scattering that has the same functional form as that of a molecular radiofrequency spectrum~\cite{Chin2005,Ketterle2008}.
At finite temperature the delta function may be naturally assumed to broaden into a Lorentzian~\cite{Mahan2000}.
The variational solution presented here turns out to be identical to  the corresponding lowest order T-matrix result~\cite{Rath2013, Li2014}. We stress that the inclusion of the non-zero boson-boson interaction strength, contributions from higher-order scattering processes, as well as finite temperature will all act to lower the quasiparticle weight $Z$ further.

\begin{figure}
		\includegraphics[width=\columnwidth]{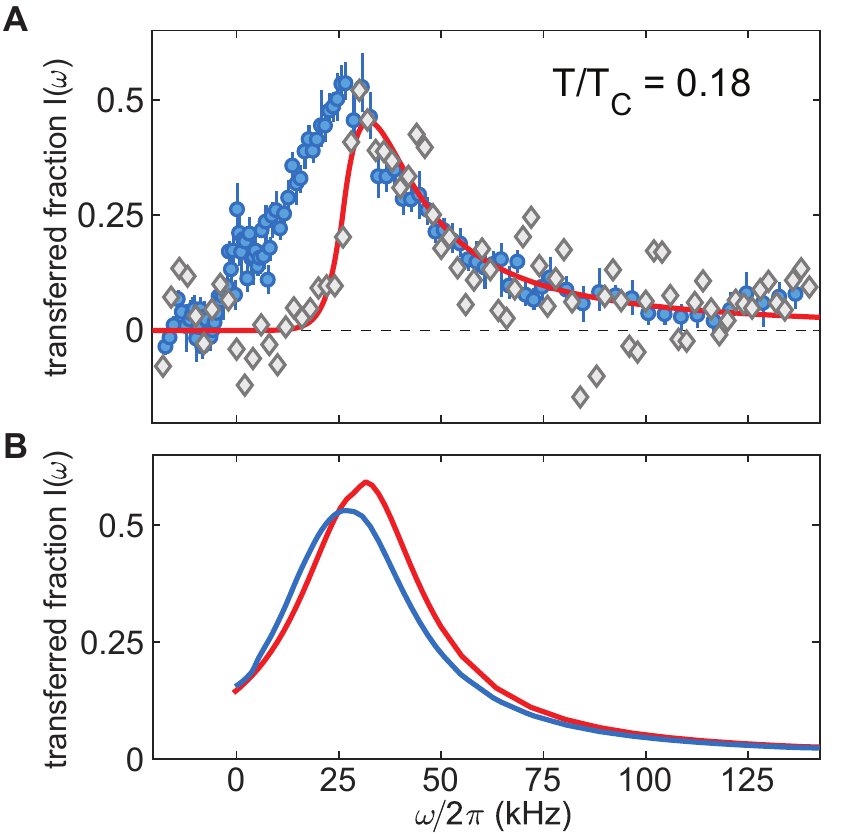}
		\caption{\label{fig:localSpec}
			Effects of column integration on the polaron lineshape. \textbf{(A)} The column-integrated rf spectrum (blue circles), as shown in the main text, is compared with the reconstructed local  rf spectrum (grey diamonds).  The red line is a guide to the eye for the local spectrum. \textbf{(B)} The variational prediction of the local rf spectrum, modeled by Eq.~\ref{eq:tmatrix}, is shown as the red line. The blue line is the same spectrum weighted by the spatial impurity distribution and integrated over the x-direction, simulating the column-integrated spectrum in (A).
		}
\end{figure}

\begin{center}
	\textbf{Homogeneous properties derived from column-integrated images}
\end{center}
\smallskip
For 3D atomic gases, absorption imaging implies line-of-sight (column) integration, resulting in a 2D projection.
We derive the local (homogeneous) properties of the Bose polaron from analyzing small regions of this 2D projection.
From the impurity absorption images, we obtain the column-integrated impurity transfer, which relates to the local transfer $I^{3\mathrm{D}}(\omega,x,y,z)$ via

\begin{equation}
I^{2\mathrm{D}}(\omega,y,z) =\frac{\int_{-\infty}^\infty \mathrm{d}x \,I^{3\mathrm{D}}(\omega,x,y,z)n_\mathrm{K}(x,y,z)}{\int_{-\infty}^\infty \mathrm{d}x \,n_\mathrm{K}(x,y,z)}
\end{equation}
where the $n_\mathrm{K}$ is the impurity density distribution.
As we will now discuss, the measured spectral peak, spectral width, and contact are only weakly affected by this column integration.
To reconstruct a local rf spectrum from an inhomogeneous sample, one can make use of the line-density-based reconstruction method described in~\cite{Ho2010,Lingham2014}.
In brief, $I^{3\mathrm{D}}(\omega,x,y,z)$ satisfies LDA and can be written as a function of the local boson chemical potential and rf frequency. In the following, we assume that the impurities share the same trap geometry as the bosons, a good approximation inside the Thomas-Fermi radius where they experience a strong attractive potential due to the BEC.
One can define a one-dimensional impurity transfer
\begin{equation}
I^{1\mathrm{D}}(\omega,z) {=} \frac{1}{n_\mathrm{K}^{1\mathrm{D}}(z)}\int\int_{-\infty}^{\infty}\mathrm{d}x\mathrm{d}y\,I^{3\mathrm{D}}(\omega,x,y,z)n_\mathrm{K}(x,y,z)
\end{equation}
where $n_\mathrm{K}^{1\mathrm{D}}(z)$ is the line density of the impurities. The one-dimensional impurity transfer and the impurity line density can be derived from the absorption images.
The local transfer follows the relation~\cite{Lingham2014}
\begin{equation}
I^{3\mathrm{D}}(\omega,x\,{=}\,y\,{=}\,0,z)=\frac{\mathrm{d}\big(I^{1\mathrm{D}}(\omega,z)n_\mathrm{K}^{1\mathrm{D}}(z)\big)}{\mathrm{d}z}\bigg(\frac{\mathrm{d}n_\mathrm{K}^{1\mathrm{D}}(z)}{\mathrm{d}z}\bigg)^{-1}
\end{equation}
Effectively, the local impurity transfer can be recovered from knowledge of the impurity line density and the doubly-integrated impurity transfer, at the cost of performing a numerical derivative.
An example of the reconstructed local spectrum is shown in Fig.~\ref{fig:localSpec}{(A)} for the axial coordinate $z\,{=}\,\frac{2}{3}R_z$, as well as the spectrum from the corresponding column-integrated data (as shown in the main text).
The discrepancy in peak energy shift between the two methods is less than 10\%.  As expected, the column-integrated spectrum has a peak shifted toward lower frequencies due to density inhomogeneity.
The spectral width of the column-integrated spectrum, as reported in the main text, is defined as the half-width-half-maximum, taking the half width toward high frequencies.  This definition of the spectral width avoids including the broadening toward lower frequencies, which is dominated by the contribution from density inhomogeneity.  The spectral width obtained from the two methods agree to within the errorbars reported in Fig. 3 of the main text.
The local reconstruction method works well near the center of the trap where the bosons and impurities have similar spatial distributions, but fails outside the BEC boundaries where the species-selective optical potential breaks the assumption of identical trap geometry between the bosons and impurities.  For this reason, we use the column-integrated transfer and not the reconstructed local transfer $I^{3\mathrm{D}}(\omega)$ for all values reported in the main text.

We perform a complementary benchmark of our column-integrated results by numerically simulating a local impurity distribution, using the result from the variational ansatz presented above, with a Lorentzian distribution centered at $E_p$ with full-width-half-maximum $\gamma$ in place of the delta-function in Eq.~\ref{eq:tmatrix}.
Upon column integration and weighted by the impurity's spatial distribution, the local spectrum gives the simulated column-integrated rf spectrum.
Here we choose $\gamma$ as a free parameter to have the best least-squares fit to our column-integrated data.
As can be seen in Fig.~\ref{fig:localSpec}{(B)}, the polaron peak position from the simulated column-integrated spectrum only deviates by approximately 10\% of the local binding energy, in agreement with the result from direct reconstruction of the local spectrum.

\begin{figure}
		\includegraphics[width=\columnwidth]{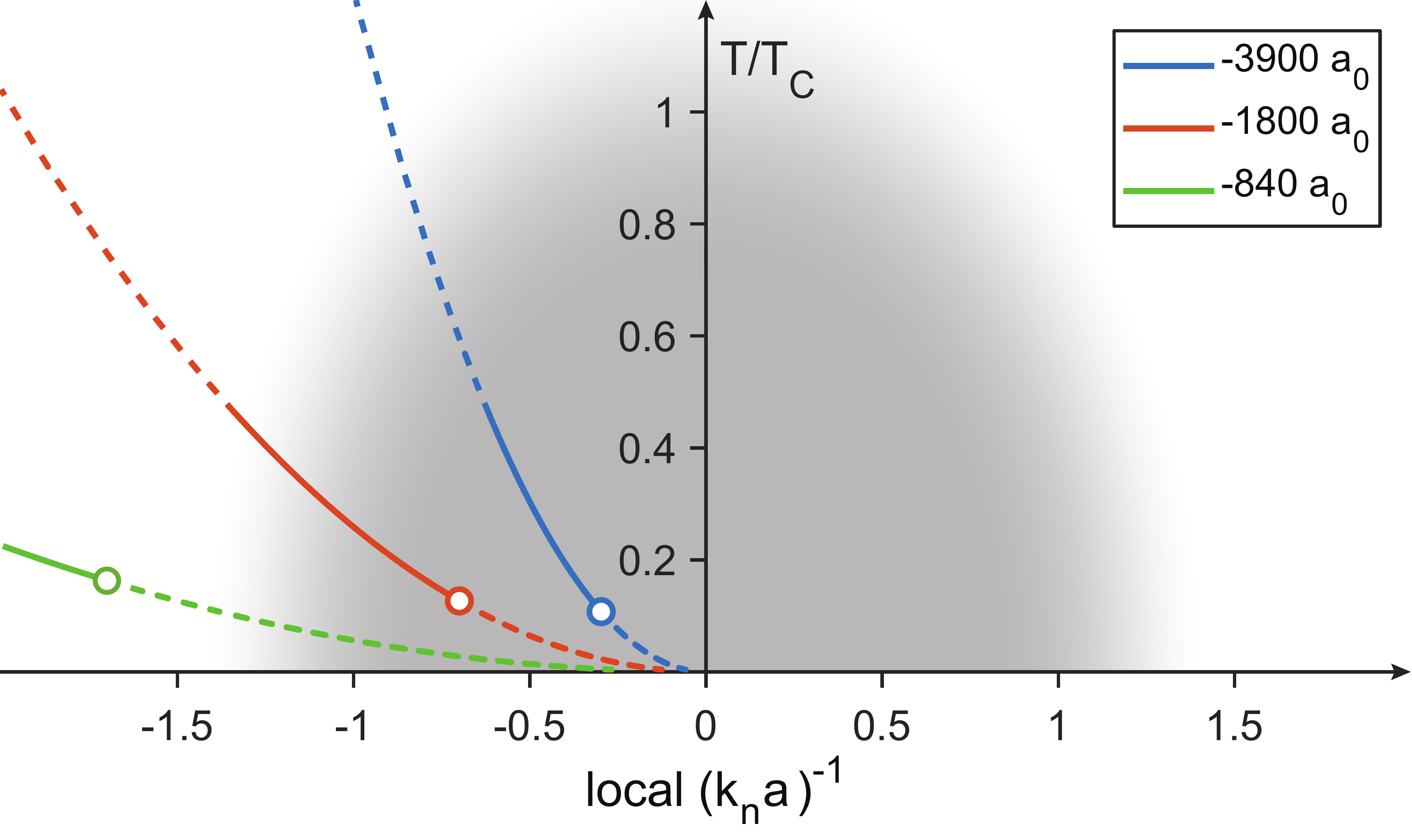}
		\caption{\label{fig:localkna}
			Relation of $T/T_\mathrm{C}$ to local $\big(k_n(z)a\big)^{-1}$ for the peak interaction strengths of $(k_na)^{-1}\,{=}\,-0.3,-0.7$, and $-1.7$ in blue, red, and green lines, respectively. Solid lines represent impurities immersed well within the BEC. Peak $(k_na)^{-1}$ as given in main text are indicated by open circles.
		}
\end{figure}

We note that in Fig.~3 of the main text, the energy and width are shown for \textit{peak} interaction strengths $(k_na)^{-1}$, but the local interaction strength $\big(k_n(z)a\big)^{-1}$ does deviate from the reported values of $-0.3,-0.7,$ and $-1.7$, due to the cloud inhomogeneity.  In Fig.~\ref{fig:localkna}, the deviation in $T/T_\mathrm{C}$ is shown as a function of local interaction strength.

\begin{figure}[tb]
	\begin{center}
		
		\includegraphics[width=\columnwidth]{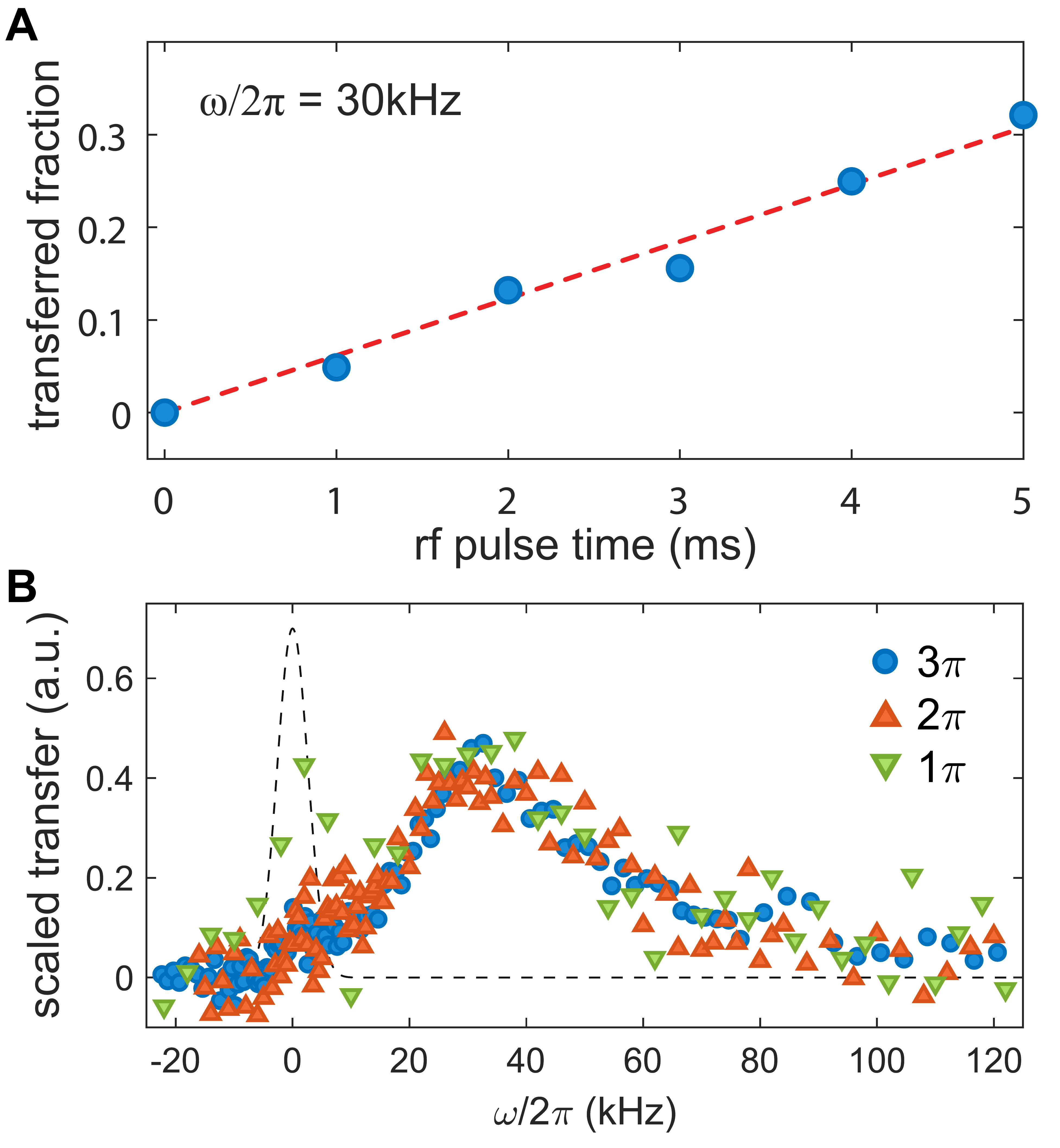}
		\caption{\label{fig:linresponse} \textbf{(A)} Time-resolved rf response of the impurity located at the trap center.  The interaction strength is $a \,{=}\, -840\,a_0$, and a rf pulse with constant power is employed with varying duration at a detuning of $2\pi \times$\SI{30}{\kilo\hertz} above the atomic resonance.  The dashed line is a linear fit through the data.
			\textbf{(B)} Rf spectra at $a\,{=}\, {-}3900\,a_0$ and $T/T_\mathrm{C}\,{=}\,  0.1$ obtained with Gaussian pulses of varying peak powers, corresponding to a $\pi,2\pi$, and $3\pi$ transfer of the bare atoms from the $\ket{\downarrow}$ to the $\ket{\uparrow}$ state.  The three powers are represented by green downward-facing triangles, red upward-facing triangles, and blue circles, respectively.
		}
	\end{center}
\end{figure}

\begin{center}
	\textbf{Linear response}
\end{center}
\smallskip
To obtain the contact from the spectral lineshape, we require the rf transfer to be in the linear response regime.  Furthermore, the contact's relation to the spectral response (Eq.~2 of the main text) is expected to only hold for the high momentum wings of the cloud, $\hbar\omega\gg E_n$.
We extract the contact by fitting the high frequency tail of the ejection spectra where the transfer is well within the linear response regime (below 0.2) and the spectral response follows a $\omega^{-3/2}$ dependence within our signal to noise, as demonstrated in the main text. To confirm the that the response is within the linear response regime, we measure the transfer as a function of the rf pulse duration.
We verify this linear behavior for various interaction strengths and Rabi frequencies and show one example in Fig.~\ref{fig:linresponse}(A).

To ensure that our polaron peak and width assignations are not affected by the possible nonlinear response of $I(\omega)$, we measure the spectrum at $a{=-}3900$ $a_0$ with varying pulse powers.  A comparison of the same spectrum taken with varying peak Rabi frequencies is shown in Fig.~\ref{fig:linresponse}(B).  The relative heights of the two lower-power spectra have been scaled by arbitrary constants.  Neither the assigned peak nor the spectral width are affected beyond the experimental signal-to-noise limitations.
\begin{center}
	\textbf{Quantum Criticality of fermionic impurities immersed in a Bose gas}
\end{center}
\smallskip
Here we discuss quantum criticality of a Bose-Fermi mixture in the highly polarized limit of impurities immersed in a Bose condensed gas. 
As pointed out in seminal works by Sachdev and Nikolic~\cite{Nikolic2007} and Sachdev~\cite{Sachdev2011}, quantum gases are generically controlled by a quantum critical point occurring at zero temperature, namely the point separating the vacuum of a given species from the phase at finite density. In the grand-canonical framework, this occurs at a particular chemical potential for that species. As with all quantum phase transitions, such as the well-known superfluid-to-Mott insulator transition with lattice bosons, the quantum critical point at zero temperature can never be experimentally reached. The point, however, controls the behavior of the system in the entire region surrounding it at non-zero temperature.
The quantum critical framework applies independently of the quantum statistics of the gases: to bosons, fermions, and their mixtures~\cite{Nikolic2007}.

A simple realization of quantum criticality in ultracold gases is found in the non-interacting Fermi gas~\cite{Sachdev2011}. The quantum critical point separates the vacuum of fermions $n_F = 0$ at $\mu_F <0$ from the Fermi liquid phase containing fermions, $n_F\sim \mu_F^{3/2}$ at $\mu_F >0$. The density in the region at non-zero temperature above this quantum critical point, $n_F \sim 1/\lambda^3$, corresponds to an interparticle distance on the order of the de Broglie wavelength $\lambda$, showing that quantum and thermal effects are equally important~\cite{Sachdev2011}.
In addition, the weakly interacting Bose gas can be discussed from the viewpoint of quantum criticality~\cite{Sachdev2011}, again with a quantum critical point at $T=0$ separating the boson vacuum $n_B=0$ at $\mu_B < 0$ from the gas at finite density $n_B = \mu_B/g_{\rm BB}$ with $g_{\rm BB} = \frac{4\pi\hbar^2 a_{\rm BB}}{m_B}$, and $a_{\rm BB}$ the Bose-Bose scattering length. The classical second-order phase transition of Bose-Einstein condensation is a line in the $T$-$\mu_B$ plane terminating at the $T=0$ quantum critical point. The region of classical criticality of critical thermal fluctuations is only a narrow sliver around that line (within the Ginzburg region), and is to be distinguished from the much wider quantum critical region at non-zero temperature and chemical potentials in the vicinity of $\mu_B=0$ at $T$ above the quantum critical point.

Turning to the Bose-Fermi mixture relevant to this work, the system is here described by four parameters, which in the grand-canonical setting are the boson and fermion chemical potentials $\mu_B$ and $\mu_F$ respectively, the coupling strength describing Bose-Bose scattering $g_{\rm BB}$, and the interspecies Bose-Fermi coupling strength $g_{\rm BF}= \frac{2\pi\hbar^2 a}{m_r}$. The ratio of fermion to boson mass $\alpha = m_F/m_B$ provides an additional parameter.
Already at zero temperature, a rich phase diagram is expected that has been the topic of intense theoretical research~\cite{Rath2013,Ludwig2011,Fratini2013,Watanabe2008}.
The richness is evident from the viewpoint of quantum criticality. The quantum {(multi-)critical} points and lines of an interacting Bose-Fermi mixture were studied in Ref.~\cite{Ludwig2011}, which distinguishes seven different phases including superfluid and Fermi liquid phases, pure vacuum, boson vacuum, and fermion vacuum. The present study focuses on the impurity limit of few fermions immersed in a Bose gas. This simplifies the situation significantly. In the presence of a weakly interacting Bose condensate with $n_B>0$, i.e. at $T \ll T_\mathrm{C}$, the quantum phase transition of primary interest is the one separating the vacuum of fermions $n_F =0$ from the Fermi liquid phase with $n_F>0$. 
These two phases are part of the seven phases described in Ref.~\cite{Ludwig2011}.

\begin{figure}
		\includegraphics[width=\columnwidth]{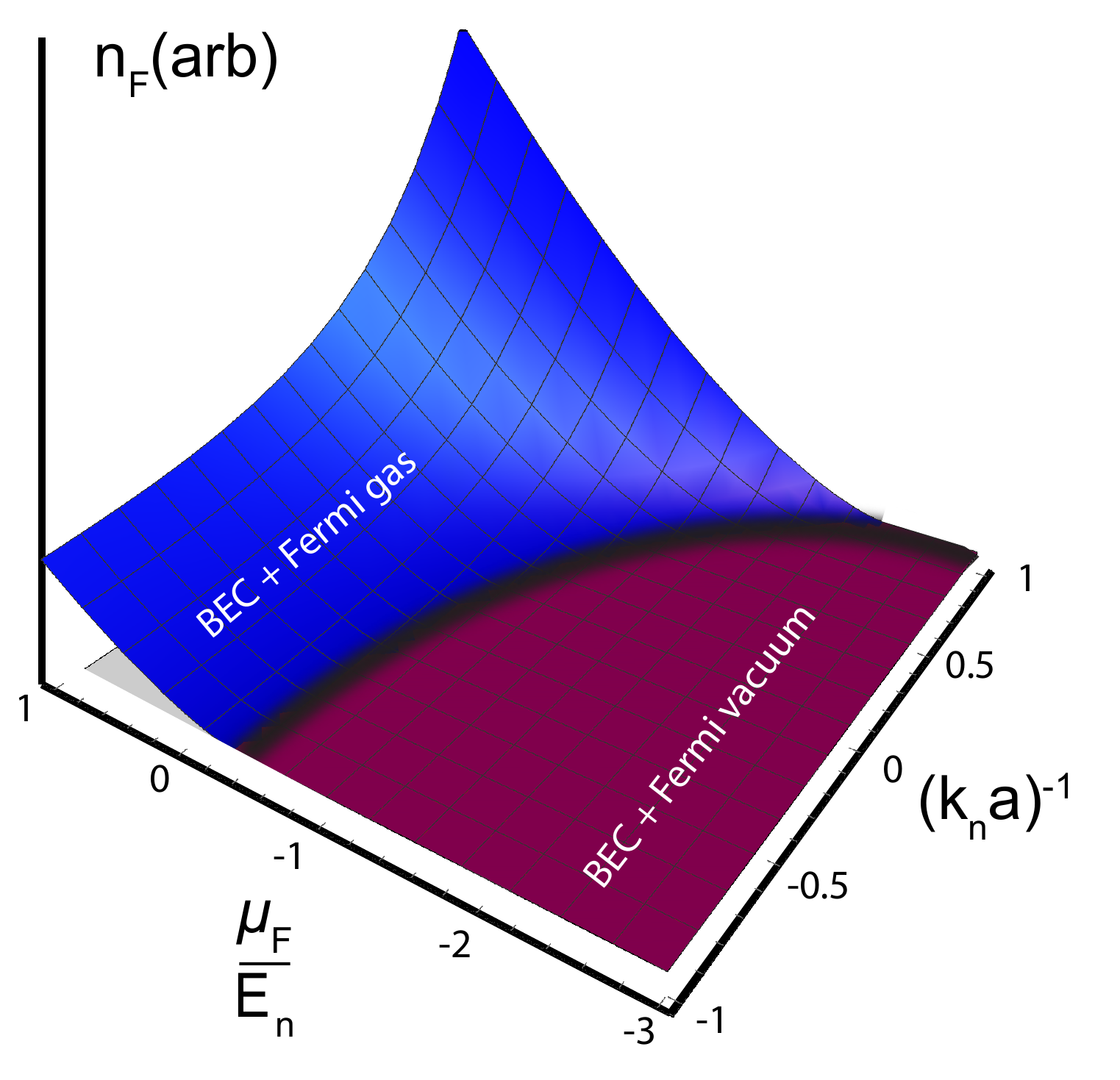}
		\caption{\label{fig:PolaronPhaseDiagram} 
			Quantum phase transition of fermions immersed in a Bose-Einstein condensate, from fermion vacuum to the phase at non-zero fermion density. The critical line is precisely given by the energy to add a single fermion to the BEC, the polaron energy. 
		}

\end{figure}
Unlike the case of non-interacting fermions, in the presence of the BEC the transition from fermion vacuum to a non-zero fermion density does not occur at $\mu_F{=}0$, but at the shifted location $\mu_F^*{=}E_p$ given by the energy of a Bose polaron. This is precisely the energy cost to add a single fermion to a Bose-Einstein condensate, and it is negative for attractive interactions.
The critical chemical potential $\mu_F^*$ that marks the quantum phase transition at $T=0$ depends on $\mu_B$, $a_{\rm BB}$ and $a$. Given that the boson density $n_B = \mu_B/g_{\rm BB}>0$ is non-zero, one can replace $a_{\rm BB}$ in favor of the energy scale $E_n = \frac{\hbar^2 k_n^2}{4 m_r}$ with $k_n = (6\pi^2 n_B)^{1/3}$. The general expression for $\mu_F^*$ is then
\begin{equation}
\mu_F^* = E_n \,f\left(\frac{1}{k_n a}, \frac{\mu_B}{E_n}\right)
\end{equation}
where $f(x,y)$ is a function of the dimensionless parameters $1/k_n a$ and  $\mu_B/E_n$.
For weakly interacting Bose gases where $n_B^{1/3}a_{\rm BB}\ll1$, however, $\mu_B \ll E_n$, and one expects the simpler relation:
\begin{equation}
\mu_F^* = E_n\, f(\frac{1}{k_n a},0)
\end{equation}
Indeed, such independence of the polaron energy on the boson chemical potential $\mu_B$ was found in studies of the Bose polaron problem within the $T$-matrix or variational approach~\cite{Rath2013,Li2014}. In the limit $\mu_B{=}0$, the variational prediction for the quantum phase transition line is given in Eq.~1 of the main text.
Fig.~\ref{fig:PolaronPhaseDiagram} shows a sketch for the $T=0$ phase diagram, with the axes $\mu_F/E_n$, $1/k_n a$ and the fermion density $n_F$.

The quantum critical line (for varying $a$) at zero temperature determines the behavior of the polaron gas at finite temperature, in close analogy with the case of a non-interacting Fermi gas. In particular, the quantum critical regime where thermal and quantum effects are both equally relevant is characterized by temperatures $T\gg \left|\mu_F - \mu_F^*\right|$. Here, as was the case for non-interacting fermions, the fermion density behaves as $n_F \sim \frac{1}{\lambda_P^3}$. The fermions are however not the bare fermions of the non-interacting theory, but dressed into polarons. Accordingly, the de Broglie wavelength of the polarons $\lambda_P{=} h/\sqrt{2\pi m_P k_B T}$ is given by the polaron mass $m_P$.

A striking difference between the impurity immersed in the Bose gas and the non-interacting Fermi gas is that scattering with the background Bose gas endows the polarons with a finite lifetime at non-zero temperature.
Again, the quantum critical viewpoint helps to estimate the behavior of the inverse lifetime of quasiparticles. 
This quantity, given by the imaginary part of the self-energy, can only depend on $T$ and $E_n$ near unitarity ($1/k_n |a| \ll 1$) and must obey the scaling relation
$\Gamma = \frac{k_B T}{\hbar} f_\Gamma (T/E_n)$, where $f_\Gamma(x)$ is a dimensionless function of a single dimensionless argument (away from unitarity, there will be additionally a dependence on $1/k_n a$, and for strong Bose-Bose interactions the ratio $\mu_B/E_n$ becomes relevant).
In the limit of low temperatures, we may expect long-lived quasiparticles (the Bose polarons) and thus $\Gamma(T\rightarrow 0) = 0$. This behavior excludes the scaling of the dimensionless function $f_\Gamma(x)\sim 1/x$ or other divergent behavior as $x\rightarrow 0$. The lowest-order possibility for the limit of $f_\Gamma(x)$ as $x\rightarrow 0$ is $f_\Gamma(x\rightarrow 0) = {\rm const.}$ The resulting linear scaling at the ``Planckian'' rate $\Gamma = {\rm const.}\, k_B T/\hbar$ is in agreement with our experimental findings, where we find ${\rm const.} \approx 8$ close to unitarity.
The very expression for $\Gamma$ involving temperature and Planck's constant reveals that quantum and thermal effects are equally important~\cite{Sachdev2011}.
An expansion of $f_\Gamma(x)$ in powers of $x$ can thus start with a non-zero constant term, and then also include additional terms $\sim x$ and $\sim x^2$ that would become relevant as $T\rightarrow E_n$.
We note that in the case of an impurity swimming in a {\it fermionic} background one has $f_\Gamma(x) \propto x$ at small $x$, so $\Gamma \propto T^2/E_n$, a direct consequence of Pauli blocking of collisions in the host Fermi gas~\cite{Yan2019}. This reminds us that the properties of the impurity are directly tied to those of the host gas. If the host gas displays quantum critical behavior on its own, this will be revealed by the impurity scattering off of the excitations in the host gas. In the present case, the impurity can thus act as a sensor also for the quantum critical region near $\mu_B=0$ of the Bose gas. Indeed, a complete description should depart from the tri-critical point found in Ref.~\cite{Ludwig2011}, separating the total particle vacuum ($\mu_B<0$, $\mu_F<\mu_F^*$), the BEC phase without fermions ($\mu_B>0$, $\mu_F<\mu_F^*$), and the BEC phase with fermions ($\mu_B>0$, $\mu_F>\mu_F^*$).
This should lead to a rich and presently theoretically largely unexplored scenario of physics at non-zero temperatures above multi-critical points and lines of a Bose-Fermi mixture.

\begin{center}
	\textbf{Estimate of the impurity decay rate}
\end{center}
\smallskip

For a direct estimate of the quasiparticle decay rate, we can look towards precise results obtained for Bose polarons in the weakly interacting regime where $1/k_n |a| \gg 1$~\cite{Levinsen2017}. There, it was found that in the presence of a BEC, at $T < T_\mathrm{C}$, the polaron decay is driven by thermally excited bosonic quasiparticles. 
The impurity thus scatters off the quantum saturated Bose gas of density $n_{\rm th} \sim \zeta(3/2) \frac{1}{\lambda_B^3}$. Here, $\lambda_B$ is the de Broglie wavelength given by the boson mass $m_B$, and $\zeta(s)$ the Riemann zeta function.
As can be seen from the results in Ref.~\cite{Levinsen2017}, the decay rate is given, up to dimensionless factors on the order of unity (that weakly depend on the Bose-Bose interaction strength $n^{1/3}a_{\rm BB}$), by
\begin{eqnarray}
\Gamma &=& n_{\rm th}\, \sigma\, v_{\rm rel}\\
&\sim& \frac{1}{\lambda_B^3} a^2 \sqrt{\frac{k_B T}{m_{\rm r}}}\\
&\sim& \left(\frac{m_B}{m_r}\right)^{3/2} \frac{m_r a^2}{\hbar^2} \frac{\left(k_B T\right)^2}{\hbar} 
\end{eqnarray}
The decay is thus quadratic in temperature. Above, $v_{\rm rel}$ is the average relative speed and $m_r$ the reduced mass.

If there is no qualitative change as the interactions increase towards unitarity, the thermal, quantum saturated boson gas will be responsible for the decay of fermionic quasiparticles. We may then replace the scattering cross section by its unitarity-limited value $\sigma \sim \lambda_{\rm rel}^2$ given by the square of the de Broglie wavelength corresponding to the particle of reduced mass $m_r$. This amounts to replacing $m_r a^2/\hbar^2$ above by $1/k_B T$, yielding the simple estimate for the polaron decay rate near unitarity
\begin{equation}
\Gamma = n_{\rm th} \, \sigma\, v_{\rm rel}  \sim \left(\frac{m_B}{m_r}\right)^{3/2} \frac{k_B T}{\hbar}
\end{equation}
\begin{figure}
		\includegraphics[width=\columnwidth]{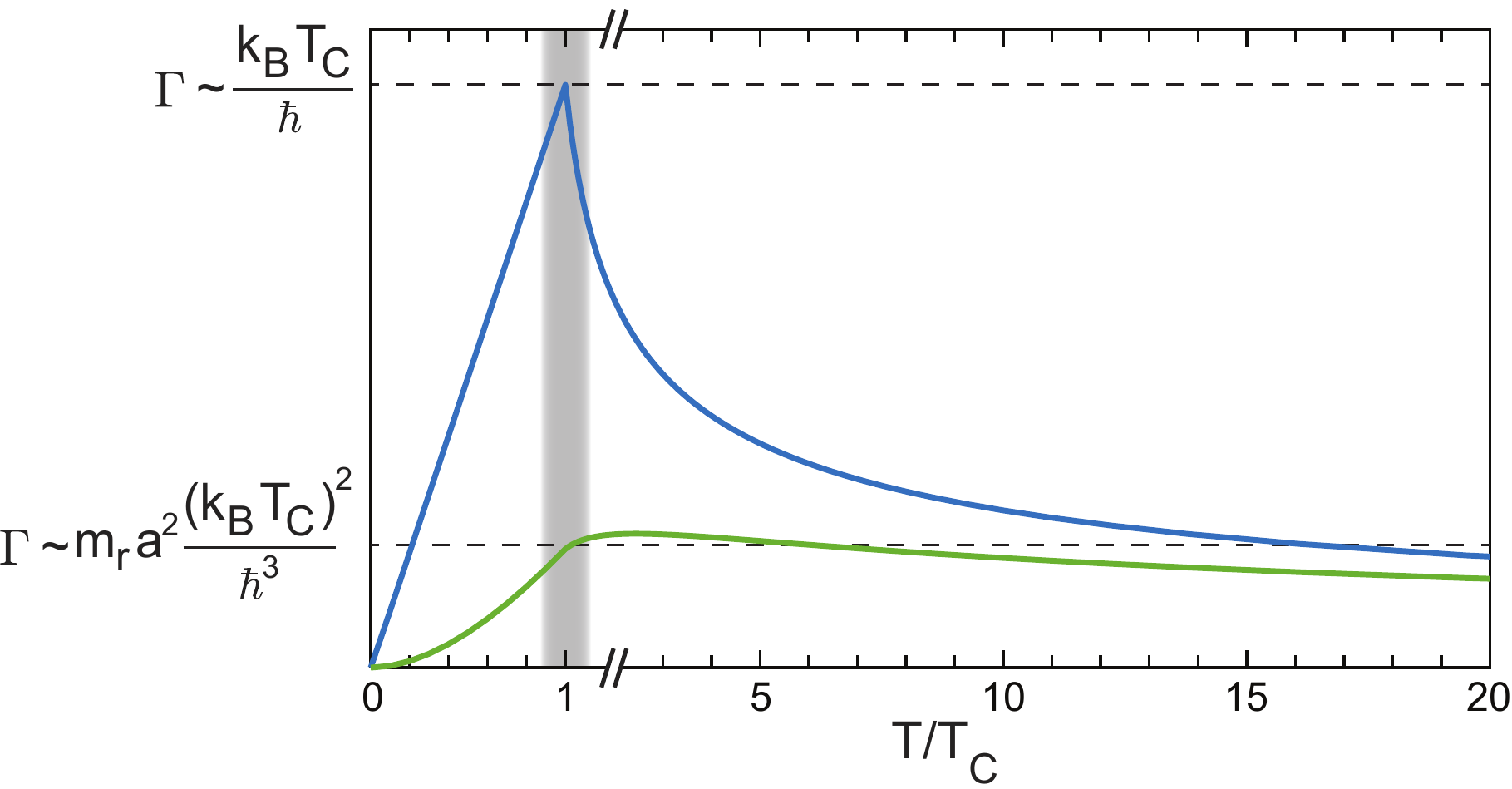}
		\caption{\label{fig:Gamma} The Bose polaron decay rate in the degenerate region $T{<}T_\mathrm{C}$ and in the Boltzmann region $T{>}T_\mathrm{C}$, for unitary interactions (blue) and non-unitary interactions (green), as obtained from the calculations in the text. In the shaded region effects due to classical critical fluctuations of the Bose gas may enter.
		}
\end{figure}
Here, $\Gamma$ is indeed a \textit{Planckian} decay rate on the scale of $k_B T/\hbar$, signaling the equal role played by quantum and thermal effects.
As $k_B T$ approaches $T_\mathrm{C}$, the scattering rate grows to $\Gamma \sim E_n/\hbar$, on the scale of the polaron energy itself. This indicates the absence of well-defined quasiparticles, another hallmark of quantum critical behavior.

In Fig.~\ref{fig:Gamma} we display the impurity scattering rate, obtained from the calculation outlined in the next section, for unitarity-limited and more weakly interacting gases.
The maximum scattering rate occurs at $T_\mathrm{C}$, and its value $\Gamma\,{\sim}\, E_n/\hbar$ is in agreement with the maximum rate observed in the experiment near unitarity.
In the above scenario, quantum depletion of the weakly interacting condensate and the regime of classical criticality of the Bose condensate in a narrow region around $T_\mathrm{C}$ (Ginzburg region) were not relevant. These aspects would need to be considered in a full description. 

\paragraph{\textit{Calculation of the impurity scattering rate}:}
Using the previous estimates as guides, we turn to a calculation of the impurity scattering rate in the thermally excited, quantum saturated Bose gas below $T_\mathrm{C}$, and in the thermal Bose gas above $T_\mathrm{C}$. For this, we take the Bose gas to be ideal, \textit{i.e.} the limit $a_{\rm BB}\,{=}\,0$ (of course thermodynamic stability requires a non-zero $a_{\rm BB}>0$). The boson momentum distribution is $f_B(\vec{k})\,{=}\, \frac{1}{e^{\beta \epsilon_{B,k}-\beta \mu_B} -1}$ with $\epsilon_{B,k} \,{=}\, \frac{\hbar^2 k^2}{2 m_B}$, where below $T_\mathrm{C}$ we have $\mu_B \,{=}\, 0$. Above $T_\mathrm{C}$ the chemical potential is related to the boson density by $n_B \lambda_B^3 = \zeta_{3/2}(e^{\beta \mu_B})$, where $\zeta_s (z)=\sum_{n=1}^\infty \frac{z^n}{n^s}$ is the Polylogarithm. The impurity as a single particle is Boltzmann-distributed in the volume $V$, so $f_F(\vec{k}) \,{=}\, \frac{\lambda_F^3}{V} e^{-\beta \epsilon_{F,k}}$ with $\lambda_F$ the de Broglie wavelength given by the fermion mass $m_F$ and $\epsilon_{F,k} \,{=}\,\frac{\hbar^2 k^2}{2 m_F}$. Below $T_\mathrm{C}$ we should replace the fermion mass by the polaron mass, but this only weakly affects the resulting decay rate. 
The average scattering rate experienced by the impurity, averaged over boson and impurity momenta, is 
\begin{eqnarray}
\Gamma &=& n_B \left<\sigma(k) v_{\rm rel}\right> \nonumber\\
&=& \frac{1}{V}\sum_{k_B} \sum_{k_F} f_B(\vec{k}_B) f_F(\vec{k}_F)\sigma(k) \frac{\hbar k}{m_r} 
\end{eqnarray}where the relative velocity $\vec{v_{\rm rel}}\,{\equiv}\, \frac{\hbar \vec{k}}{m_r}\,{=}\,\frac{\hbar \vec{k_F}}{m_F}{-}\frac{\hbar \vec{k_B}}{m_B}$ is the difference between the boson and fermion velocities, and $\hbar \vec{k}$ is the relative momentum of the scattering particles. The scattering cross section is, to lowest order, the one for two particles interacting in vacuum, $\sigma(k) = 4\pi a^2/(1+k^2 a^2)$. Writing the Bose distribution function as an infinite series, we have

\begin{eqnarray}
\Gamma &=& \lambda_F^3 \sum_{n=1}^\infty e^{n \beta\mu_B} \int \frac{{\rm d}^3k_B}{(2\pi)^3} \int \frac{{\rm d}^3k_F}{(2\pi)^3}\times \nonumber\\
&&\left(e^{-n \,k_B^2 \lambda_B^2/4\pi} e^{-k_F^2 \lambda_F^2/4\pi} \frac{4\pi a^2}{1 + k^2 a^2} \frac{\hbar k}{m_r}\right)
\end{eqnarray}
Introducing a wave vector $\vec{K} \,{=}\, \frac{m_n}{m_r} (\vec{k_B} + \frac{1}{n}\vec{k_F})$ with $m_n \,{=}\, \frac{n\, m_B m_F}{m_F + n \,m_B}$ the reduced mass of a fermion $m_F$ and a particle consisting of $n$ bosons, we transform from $\vec{k_B}$ and $\vec{k_F}$ to $\vec{k}$ and $\vec{K}$ and perform the integrations.
The result is
\begin{equation}
\Gamma= \frac{1}{\lambda_B^3} 4\pi a^2 \,v_{\rm rel}\,  g\left(\frac{T}{T_a},z,\alpha\right)
\end{equation}
where $g$ is a dimensionless function that only depends on the ratio $T/T_a$, with $T_a \,{=}\, \hbar^2/(2 m_r a^2 k_B)$ the temperature scale associated with the scattering length, the fugacity $z \,{=}\, e^{\beta\mu_B}$ of the bosons, and the mass ratio $\alpha \,{=}\, m_F/m_B$. Explicitly, $g$ can be written as
\begin{equation}
4\pi a^2 g\left(\frac{T}{T_a}, z, \alpha\right) = \sum_{n=1}^\infty \frac{z^{n}}{n^{3/2}}\sqrt{\frac{m_r}{m_n}} \sigma\left(\frac{T}{T_a} \frac{m_r}{m_n}\right)
\end{equation}
where $\frac{m_r}{m_n} \,{=}\, \frac{1}{n}\frac{\alpha+n}{\alpha + 1}$ and
\begin{equation}
\sigma\left(\frac{T}{T_a}\right) = 8\pi a^2 \int {\rm d}x\, x^3 \frac{1}{1+ \frac{T}{T_a} x^2} e^{-x^2}
\end{equation}
where $\sigma$ is a thermally averaged scattering cross section, interpolating between $\sigma(T\ll T_a) \,{=}\, \allowbreak 4\pi a^2$ and $\sigma(T\gg T_a) \,{=}\, 4\pi a^2 T_a/T$. For near-resonant interactions one has $\sigma(T\gg T_a) \,{=}\, \lambda_{\rm rel}^2$. A similar (but not identical) thermally averaged cross section is found for spin transport~\cite{Sommer2011}.
We can now discuss the impurity decay rate in the classical ($T>T_\mathrm{C}$) and the quantum degenerate limit ($T<T_\mathrm{C}$), and for weak and resonant interactions. First, in the classical regime $z \ll 1$ at weak interactions $T\ll T_a$, we retrieve the well-known result $\Gamma = n_B\, 4\pi a^2\, v_{\rm rel}$ which increases with temperature like $\sqrt{T}$. In the limit $T\gg T_a$ however, we enter the realm of the unitary Boltzmann gas, where we have
\begin{equation}
\Gamma_{\rm Boltzmann} {=}  n_B \lambda_{\rm rel}^2 v_{\rm rel} {=} \frac{16\sqrt{2}}{3\pi^{3/2}}\frac{E_n}{\hbar} \sqrt{\frac{E_n}{T}} {\approx} 1.35 \frac{E_n}{\hbar} \sqrt{\frac{E_n}{T}} 
\end{equation}
So the scattering rate decreases as $1/\sqrt{T}$ and one has bare, undressed fermions at high temperatures $T\gg \{T_a, E_n\}$~\cite{Yan2019}. The result is also obtained from the thermal average of the decay rate $\Gamma(\vec{p})\,{=}\, -2\, {\rm Im} \Sigma(\epsilon_p,\vec{p})$ for an impurity of momentum $\vec{p}$ using the high-temperature expression for the self-energy~\cite{Enss2011}.
As we enter the quantum degenerate regime of the Bose gas, $z\le 1$, we have for $T\,{\ll}\,T_a$ (weak interactions) $\Gamma \,{=}\, \frac{g(0,z,\alpha)}{\zeta_{3/2}(z)} \, n_B \,4\pi a^2\, v_{\rm rel}$. The numerical coefficient $\frac{g(0,z,\alpha)}{\zeta_{3/2}(z)} {\rightarrow} 1$ in the limit of light impurities $\alpha {\ll} 1$, as then the relative velocity is always given by the impurity velocity. In the limit of heavy impurities $\alpha {\gg} 1$, the coefficient decreases at most to $\zeta_2(1)/\zeta_{3/2}(1) \,{=}\, 0.63$ at $z\,{=}\,1$. Compared to the Boltzmann gas, the quantum degenerate Bose gas (near, but above $T_\mathrm{C}$) has more bosons at low momenta, and therefore a reduced average velocity. For $\alpha = 40/23$, the case of $^{40}$K immersed in $^{23}$Na, one has $\frac{g(0,1,\alpha)}{\zeta_{3/2}(1)} \,{=}\, 0.82$.
For the resonant case $T\gg T_a$ and $z\le 1$, we find $\Gamma \,{=}\, \frac{h(z,\alpha)}{\zeta_{3/2}(z)}\,n_B \lambda_{\rm rel}^2 v_{\rm rel}$, similar to the Boltzmann regime, with $h(z,\alpha)\,{=}\, \sum_{n=1}^\infty \frac{z^n}{n}\sqrt{\frac{\alpha+1}{\alpha+n}}$. As before, $\frac{h(z,\alpha)}{\zeta_{3/2}(z)}$ is a numerical coefficient close to unity for all relevant mass ratios, with its maximum at $z\,{=}\,1$ growing from $\frac{h(z,\alpha)}{\zeta_{3/2}(z)}{\rightarrow} 1$ at $\alpha \,{\ll}\, 1$ to 2.53 for $\alpha \,{=}\, 100$. The reason for the increase in $\Gamma$ upon approach of the degenerate regime is again the increased number of low-momentum bosons compared to the Boltzmann regime, given that for resonant interactions $\Gamma$ is the thermal average of $n_B\left<\sigma(k) \hbar k/m_r\right> \sim \left<\frac{1}{k}\right>$, the inverse relative momentum, and is no longer $\sim \left<k\right>$, the relative momentum average, required in the weakly interacting regime. For the same reason, formally, the coefficient $h(1,\alpha)$ does eventually diverge logarithmically as $\alpha \rightarrow \infty$. This divergence of the average scattering rate when only bosons are mobile is due to the large fraction of low-momentum bosons $f_B(k\rightarrow 0) \sim T/\epsilon_k$ in the non-interacting Bose gas. The unphysical divergence is not present in interacting Bose gases, where $f_B(k\rightarrow 0) \sim T/\hbar c k$, with $c$ the speed of sound. Then, $\Gamma \sim \frac{\hbar}{m_B} \int_0^{m_B c/\hbar} {\rm d}k\, T/\hbar c \sim k_B T/\hbar$, which, naturally, is the expectation from quantum criticality.

Below $T_\mathrm{C}$, at $z=1$, the thermal Bose gas is quantum saturated, with $n_B = \zeta(3/2)/\lambda_B^3$. For weak interactions we then find a scattering rate
\begin{equation}
\Gamma {=} g(0,1,\alpha)\,\frac{1}{\lambda_B^3} 4\pi a^2 v_{\rm rel}  {=}g(0,1,\alpha) \frac{2}{\pi}  \left(\frac{m_B}{m_r}\right)^{3/2} \frac{k_B T^2}{\hbar T_a},
\end{equation}
quadratic in temperature, in agreement with the estimate above. The numerical prefactor \allowbreak $g(0,1,\alpha)=\sum_{n=1}^\infty \frac{1}{n^2} \sqrt{\frac{\alpha+n}{\alpha+1}}$ varies little, from $\zeta(3/2)\,{\approx}\,2.612$ for light fermions to $\zeta(2) \,{\approx}\, 1.645 $ for infinitely heavy fermions.
For resonant interactions below $T_\mathrm{C}$, we instead find

\begin{eqnarray}
\Gamma &=& h(1,\alpha) \frac{1}{\lambda_B^3} \lambda_{\rm rel}^2 v_{\rm rel} \nonumber\\
&=& h(1,\alpha) \frac{2}{\pi} \left(\frac{m_B}{m_r}\right)^{3/2} \frac{k_B T}{\hbar}
\end{eqnarray}
This indeed has the form of the scattering rate expected from quantum criticality.
For the present Bose-Fermi mixture of $^{40}$K immersed in $^{23}$Na, we find $\Gamma \,{\approx}\, 4.2 \frac{k_B T}{\hbar}$. If instead one uses the polaron mass, \textit{e.g.} from the variational / T-matrix calculation~\cite{Rath2013}, the prefactor is 4.0. Despite the explicit dependence on the mass ratio $\alpha$, the prefactor depends only weakly on $\alpha$.  For $\alpha\,{=}\,$1, 100, and 1000, we obtain $h(1,\alpha) \frac{2}{\pi} \left(\frac{m_B}{m_r}\right)^{3/2}\,{=}\, 5.6, 4.3$, and 5.7. 
The logarithmic divergence of $h(1,\alpha)$ at large $\alpha$ was discussed above.
For $\alpha \ll 1$ the prefactor diverges with the reduced mass like $\left(\frac{m_B}{m_r}\right)^{3/2}$. Indeed, the maximum scattering rate is reached at $T_\mathrm{C}$, where it is
\begin{equation}
\Gamma_{\rm max} = \frac{h(1,\alpha)}{\zeta(3/2)} \frac{16\sqrt{2}}{3\pi^{3/2}}\sqrt{\frac{E_n}{T_\mathrm{C}}} \frac{E_n}{\hbar}.
\end{equation}
This maximum sets the magnitude of the linear slope below $T_\mathrm{C}$. When expressed in terms of $E_n$, the maximum scattering rate is only weakly dependent on the mass ratio, $\Gamma_{\rm max}/E_n = $ 4.9, 2.4 and 3.7 for $\alpha = 0.1$, 1 and 100. We see that physically, it is the scale of $E_n$, containing the reduced mass, and not of $T_\mathrm{C}$, that governs the magnitude of the scattering rate. Indeed, given a mean-free path $l$ as short as possible for contact interactions, one interboson distance,
and an impurity velocity $v \sim \hbar/(m_r l)$ given by the Heisenberg uncertainty relation from the position uncertainty $l$, the scattering rate naturally becomes $\Gamma \sim v/l = E_n/\hbar$.

\end{document}